\documentclass[12pt]{article}
\usepackage{amsfonts,amsthm,amsmath,amssymb,upgreek,bm}
\usepackage{hyperref}
\usepackage[paper=letterpaper,margin=.85in]{geometry}
\usepackage{graphicx}
\usepackage{units}
\usepackage{float}
\usepackage[export]{adjustbox}
\usepackage{bbold}
\usepackage{xcolor}
\usepackage[shortlabels]{enumitem}
\usepackage{dsfont}
\usepackage[mathscr]{euscript}
\usepackage[normalem]{ulem}

\newcommand{\be}{\begin{equation}}
\newcommand{\ee}{\end{equation}}

\renewcommand{\i}{\mathrm{i}}
\renewcommand{\d}{\mathrm{d}}
\newcommand{\D}{\mathcal{D}}

\newcommand{\R}{\mathsf{R}}

\newcommand{\Z}{\mathscr{Z}}
\newcommand{\Zs}{\pmb{\mathscr{Z}}}
\renewcommand{\L}{L}
\renewcommand{\R}{R}
\newcommand{\LL}{L'}
\newcommand{\RR}{R'}

\newcommand{\lr}{\langle}

\newcommand{\rr}{\rangle}

\numberwithin{equation}{section}

\begin{document}
\thispagestyle{empty}

\vspace*{2.5cm}
\begin{center}

{\bf {\LARGE Wormholes without averaging}}

\begin{center}

\vspace{1cm}

 {\bf Phil Saad$^1$, Stephen H. Shenker$^2$, Douglas Stanford$^2$, and Shunyu Yao$^2$}\\
  \bigskip \rm
  
\bigskip
$^1$\hspace{.05em}School of Natural Sciences,\\ Institute for Advanced Study, Princeton, NJ 08540

\bigskip 

$^2$\hspace{.05em}Stanford Institute for Theoretical Physics,\\Stanford University, Stanford, CA 94305

\rm
  \end{center}

\vspace{2.5cm}
{\bf Abstract}
\end{center}
\begin{quotation}
\noindent

After averaging over fermion couplings, SYK has a collective field description that sometimes has ``wormhole" solutions. We study the fate of these wormholes when the couplings are  fixed. Working mainly in a simple model, we find that the wormhole saddles persist, but that new saddles also appear elsewhere in the integration space -- ``half-wormholes.''  The wormhole contributions depend only weakly on the specific choice of couplings, while the half-wormhole contributions are strongly sensitive. The half-wormholes are crucial for factorization of decoupled systems with fixed couplings, but they vanish after averaging, leaving the non-factorizing wormhole behind.

\end{quotation}

\setcounter{page}{0}
\setcounter{tocdepth}{3}
\setcounter{footnote}{0}
\newpage

\setcounter{page}{2}
\tableofcontents

\section{Introduction}
\subsection{General remarks}
Spacetime wormholes have played a significant role in recent progress  in understanding the  nonperturbative physics of quantum black holes.    Examples include: the eternal traversable wormhole \cite{Maldacena:2018lmt};  the long-time behavior of the spectral form factor \cite{Saad:2018bqo,Saad:2019lba} and correlation functions \cite{Saad:2019pqd}; and the Page curve \cite{Almheiri:2019qdq,Penington:2019kki} and squared matrix elements \cite{Stanford:2020wkf} in models of evaporating black holes.

But wormholes also lead to puzzles, in particular the factorization problem \cite{Maldacena:2004rf}.   Imagine two decoupled boundary systems in the AdS/CFT context,  labelled $L$ and $R$.  From the boundary perspective, if one evaluates a partition function in the combined system the result is just the product of the results for the two component systems: $Z_{LR} = Z_L Z_R$.  It factorizes.   But if the bulk calculation of $Z_{LR}$ includes a wormhole linking $L$ and $R$ then superficially at least $Z_{LR} \neq Z_L Z_R$.  It fails to factorize. Some of the phenomena recently explained by wormholes, in particular the spectral form factor and squared matrix elements, are described by decoupled boundary systems and so the wormhole explanation give rise to a factorization puzzle.\footnote{In the introduction we are using $Z_L, Z_R$ and $Z_{LR}$ in a notation that is hopefully self-explanatory. In the rest of the paper we will need to use more precise notation, collected in a glossary in appendix \ref{app:glossary}.}
  
The most controlled calculations of these factorizing quantities have been done in systems like the SYK model  \cite{Sachdev:1992fk,KitaevTalks,Kitaev:2017awl} and its low-energy limit JT gravity \cite{Jackiw:1984je,Teitelboim:1983ux,almheiri2015models,Maldacena:2016hyu,Jensen:2016pah,Maldacena:2016upp,Engelsoy:2016xyb} that are dual to an ensemble of boundary quantum systems \cite{Saad:2019lba,Stanford:2019vob}.  Averaging the L and R systems over the same ensemble, denoted by $\langle \cdot \rangle$,  removes the obvious factorization puzzle because $\langle Z_L Z_R \rangle$ need not be the same as $\langle Z_L\rangle\langle Z_R \rangle$.   And in fact the link between wormholes and ensembles  is an old one, going back to the ideas of Coleman \cite{Coleman:1988cy} and Giddings-Strominger \cite{Giddings:1988cx} in the 1980s.  These ideas have recently been  recast in the AdS/CFT context, with important extensions, by Marolf and Maxfield \cite{Marolf:2020xie}.\footnote{For a sampling of additional recent work on the connection between wormholes and ensembles see also \cite{Pollack:2020gfa,Afkhami-Jeddi:2020ezh,Maloney:2020nni,Belin:2020hea,Cotler:2020ugk,Bousso:2020kmy,Anous:2020lka,Chen:2020tes,Liu:2020jsv,Langhoff:2020jqa,Marolf:2021kjc,Casali:2021ewu,Meruliya:2021utr,Eberhardt:2021jvj}.}
 
We can formulate a version of the factorization puzzle in such ensembles by asking what happens to the wormholes connecting decoupled systems when we focus on just one element of the ensemble.  This question has been addressed in the Marolf-Maxfield model and  in JT gravity \cite{Blommaert:2019wfy,Marolf:2020xie,SSYtoappear,Saadtalk}.  In this paper we will examine it in the SYK model where instead of averaging we will choose a {\it  fixed} set of couplings between the fermions.

At large $N$, the SYK model can be studied using collective fields called $G$ and $\Sigma$. In some respects, these are similar to the bulk description of a holographic theory. 
Averaged single replica quantities like $\langle Z_L \rangle$ have such a ``bulk" description in terms of the collective fields we will refer to as  $G_{LL}, \Sigma_{LL}$ where these fields describe correlations within the single $L$ system.   (We will often not refer to the $\Sigma$ fields explicitly, but it should be understood that they accompany each $G$.)

To compute averaged two-replica quantities like $\langle Z_L Z_R\rangle$, one has to introduce a matrix of these $G,\Sigma$ variables, where the diagonal entries like $G_{LL}$ and $G_{RR}$ represent correlations within a given replica, and the off-diagonal $G_{LR}$ represents correlation between the replicas.  We will view this  $G, \Sigma$ action as a proxy for the bulk calculation of $\langle Z_{LR }\rangle$ and refer to a saddle point of the action with nonzero $G_{LR}$  as a wormhole:
\be
G_{LR} \neq 0 \hspace{20pt} \longleftrightarrow \hspace{20pt} \text{wormhole}.
\ee

Typically, this $G,\Sigma$ formalism is used to describe the SYK model after averaging over couplings. But it turns out, as we will see, that one can still use these same variables with fixed couplings.\footnote{We thank Milind Shyani for a thought-provoking question about this point.} The new ingredient is that the action is now a more complicated function that depends on the values of those couplings.  So we can now use the collective field formalism to study the factorization puzzle.\footnote{Collective fields and a Hilbert space factorization problem have been discussed in \cite{Harlow:2015lma,Harlow:2018tqv}.}

\subsection{Summary of results}
We are able to analyze fixed-coupling partition functions in some detail in very simple reductions of the SYK model where we restrict the time contour to one or two time points. For a single time point $Z_L$ or $Z_R$ is described by a finite dimensional integral over $N$ Grassmann variables whose action involves the random couplings $J$. Our main findings are as follows:

\begin{enumerate}
\item {\bf Wormholes persist} \\
The wormhole saddle point found in the collective field description of the ensemble averaged quantity $\langle Z_{LR} \rangle$
continues to exist for fixed couplings, and the dependence on the couplings is weak.   The wormhole is ``self-averaging."
\item {\bf Half-wormholes exist} \\
New saddles appear in the computation of $Z_L$ and $Z_{LR}$ for fixed couplings, which we interpret as ``half-wormholes."  The contribution of half-wormholes depends sensitively on the particular choice of couplings -- they are ``non-self-averaging."  Together with disk and wormhole saddles, these give an accurate evaluation of $Z_L$ or $Z_{LR}$.
\item {\bf Multiple bulk descriptions coexist}\\
Without averaging, there is a choice of collective field (bulk) description for $Z_{LR}$. One can use a description that includes $G_{LR}$, or one can compute $Z_L$ and $Z_R$ by separate manifestly factorized computations with no $G_{LR}$ variable. The equivalence between them follows from an identity
\be
1 = \int \d G_{LR}  \ \delta(G_{LR} - \text{fermion bilinear}) e^{f(G_{LR}) - f(\text{fermion bilinear})}.
\ee
This looks like a rather trivial statement. However, after introducing an integral representation of the $\delta$ function and then approximating the integrals semiclassically, the equivalence to the manifestly factorized approach is no longer obvious. It implies that the half-wormhole saddle points in the manifestly factorized approach account for both the wormhole {\it and} half-wormhole saddle points in the approach with $G_{LR}$.  So in the description including wormholes it is the half-wormholes that restore factorization.   After averaging the half-wormholes disappear, leaving only the non-factorizing wormhole.
 \end{enumerate}

We expect, making certain assumptions we find plausible,  that this structure also describes the full SYK model.

\section{SYK with one time point}\label{sectiononepoint}
In this section, we will study the following integral over $N$ Grassmann numbers $\psi_1,\dots \psi_N$
\be\label{defz}
z =  \int \d^N\psi \exp\bigg\{\i^{q/2}\hspace{-15pt}\sum_{1\le i_1<\dots<i_q \le N}J_{i_1\dots i_q}\psi_{i_1\dots i_q}\bigg\}, \hspace{20pt} \psi_{i_1\dots i_q}\equiv \psi_{i_1}\psi_{i_2}\dots \psi_{i_q}.
\ee
The quantity $z$ can be thought of as a version of the SYK partition function where we replace the time contour by a single instant of time.\footnote{For the most part we take $q$ to be an even number greater than two. The case $q = 2$ is different and is studied in appendix \ref{appendixqequaltwo}. Aspects of the model with two time points were studied in \cite{Cotler:2016fpe}.} This system is simple enough that we will be able to analyze it in detail.

As in the SYK model, computations simplify if we consider averages over the $J$ tensor, with respect to a Gaussian distribution such that
\be\label{Jdist}
\langle J_{i_1\dots i_q}\rangle = 0, \hspace{20pt} \langle J_{i_1\dots i_q}J_{j_1\dots j_q}\rangle = \frac{(q-1)!}{N^{q-1}} \delta_{i_1j_1}\dots\delta_{i_qj_q}.
\ee
We will begin by studying the averaged theory, and then turn to our real interest, which is the theory with fixed couplings.

\subsection{Averaged theory}\label{subsectionaveragedtheory}
\subsubsection{Computing $\langle z^2\rangle$} 
The average of $z$ over the ensemble (\ref{Jdist}) vanishes, $\langle z\rangle = 0$. This means that $\langle z^2\rangle$ is the simplest nontrivial averaged quantity. To compute it, one can use a version of the ``$G,\Sigma$'' collective field formalism that is often used in studies of the SYK model \cite{Sachdev:1992fk,KitaevTalks}. To derive this, we start by writing the formula for $z^2$:
\begin{align}\label{z2defn}
z^2 =z_Lz_R&= \int \d^{N}\psi^L\d^N\psi^R \exp\bigg\{\i^{q/2}\hspace{-15pt}\sum_{1\le i_1<\dots<i_1\le N}J_{i_1\dots i_q}\left(\psi_{i_1\dots i_q}^\L + \psi_{i_1\dots i_q}^\R\right)\bigg\}.
\end{align}
After averaging over $J$ with a Gaussian distribution satisfying (\ref{Jdist}) and using that the square of a Grassmann vanishes, this becomes
\begin{align}
\langle z^2\rangle &= \int \d^{2N}\psi \exp\bigg\{\frac{(q-1)!}{N^{q-1}}\sum_{1\le i_1<\dots<i_1\le N}\psi_{i_1}^\L\psi_{i_1}^\R\dots \psi_{i_q}^\L\psi_{i_q}^\R\bigg\}\\
&= \int \d^{2N}\psi\exp\bigg\{\frac{N}{q}\Big(\tfrac{1}{N}\sum_{i = 1}^N \psi_i^\L\psi_i^\R\Big)^q\bigg\}.
\end{align}
We now introduce the basic $G,\Sigma$ trick that will be used frequently below. 
\begin{align}
\langle z^2\rangle &= \int \d^{2N}\psi \int_{\mathbb{R}} \d G \ \delta\Big(G-\tfrac{1}{N}\sum_{i = 1}^N \psi_i^\L\psi_i^\R\Big) \exp\bigg\{\frac{N}{q}G^q\bigg\}\\ \label{introduceDelta}
&= \int \d^{2N}\psi \int_{\mathbb{R}} \d G \int_{\i\mathbb{R}}\frac{\d \Sigma}{2\pi \i/N}\exp\bigg\{-\Sigma\Big(NG-\sum_{i = 1}^N \psi_i^\L\psi_i^\R\Big)\bigg\}\exp\bigg\{\frac{N}{q}G^q\bigg\}\\
&=\int_{\mathbb{R}} \d G \int_{\i\mathbb{R}}\frac{\d \Sigma}{2\pi \i/N}\exp\bigg\{N\left(\log(\Sigma) - \Sigma G + \frac{1}{q} G^q\right)\bigg\}.\label{zsquaredintrep1}
\end{align}
In the second line, the integral over $\Sigma$ was introduced as an integral representation of the delta function. In the final line, we integrated out the fermions. In a more general computation, one would have introduced a matrix of collective fields $G_{LL},G_{LR},G_{RR}$. However, for this simple model, only the $G_{LR}$ field is necesary, and to reduce clutter we will omit the subscript. The reader should keep in mind that $G$ is a $G_{LR}$ variable, expressing ``wormhole-type'' correlation.

How well-defined is this integral? Initially, we define it by integrating $\Sigma$ over the imaginary axis and $G$ over the real axis. To get a convergent answer it is important to do the integral over $\Sigma$ before the integral over $G$. The integral over $\Sigma$ is then of the form $\int_{\i\mathbb{R}} \d\Sigma \Sigma^N e^{-N\Sigma G} \propto \partial_{G}^N \delta(G)$, and the resulting $G$ integral is
\begin{align}\label{z2deriv}
\langle z^2\rangle &= N^{-N}\int_{\mathbb{R}} \d G e^{\frac{N}{q}G^q}\left(-\partial_G\right)^N\delta(G)\\
&= N^{-N} (\partial_G)^N e^{\frac{N}{q}G^q}\Big|_{G = 0} \label{Deriv1}\\
&= \frac{N!(N/q)^{N/q}}{N^N (N/q)!} \hspace{20pt} (\text{assuming $N$ is a multiple of $q$}) \label{exactans1}\\
&\approx \sqrt{q}e^{-(1 - \frac{1}{q})N} \hspace{20pt} (\text{for large $N$}).\label{zsquaredans}
\end{align}
(If $N$ is not a multiple of $q$, the answer for the integral is zero.) In this way of thinking about the integral, the $G$ variable is localized to a neighborhood of the origin. This is consistent with the fact that $G$ is supposed to represent a bilinear of fermions, which can be thought of as infinitesimal.

However, there is another way of thinking about the integral that breathes a bit more life into the $G$ variable. Without changing the answer for the integral, we can rotate the contours by defining
\be\label{rotContour}
\Sigma = \i e^{-\i\phi}\sigma, \hspace{20pt} G = e^{\i\phi }g
\ee
and integrating (first) over real $\sigma$ and then over real $g$. When $\phi = 0$, this is the contour we started with, but when $\phi = \frac{\pi}{q}$ we end up with
\be\label{zsquaredintrep2}
\langle z^2\rangle = \int_{\mathbb{R}} \d g\int_{\mathbb{R}}\frac{\d\sigma}{2\pi/N} \exp\left\{N\left(\log(\i e^{-\frac{\i\pi}{q}}\sigma) - \i \sigma g - \frac{1}{q}g^q\right)\right\}.
\ee
In this form the integral is convergent for either order of integration.

Let's see how to get the large $N$ answer (\ref{zsquaredans}) from a saddle point approximation to this integral. The saddle point equations are
\be
\frac{1}{\sigma}-\i g = 0, \hspace{20pt} -\i \sigma - g^{q-1} = 0 \hspace{20pt} \implies\hspace{20pt}  g^q = -1.
\ee
There are $q$ solutions. The real part of the on-shell action is the same for each of the solutions, but they contribute with different phases. Including the one-loop determinant, one finds that the different saddles contribute
\be
\frac{1}{\sqrt{q}} e^{-(1-\frac{1}{q})N}e^{2\pi \i m N/q}, \hspace{20pt} m = 0,\dots,q-1.
\ee
Which of these are we supposed to include? The correct answer is that we should sum over all of them. The most general-purpose way to justify this is to show that the Lefschetz antithimbles of each of the saddle points intersect with the defining contour. We have checked this by numerically solving the upward flow equations that define the antithimbles. A second way that reduces to one-dimensional contour integration logic is shown in Appendix \ref{Psiapp}. A third way is to observe that by summing over these saddle points we reproduce the large $N$ limit of the exact answer (\ref{zsquaredans}), and in particular we reproduce the fact that the answer is zero if $N$ is not a multiple of $q$.

Note that these saddle points have a nonzero value for the variable $G$, which is fixed by $\Sigma$ to be equal to $\frac{1}{N}\sum_i \psi_i^\L\psi_i^\R$. This is a ``$G_{LR}$'' variable that expresses correlation between two decoupled partition functions. So in particular, the saddle points with $G \neq 0$ represent correlation between the two replicas, analogous to a wormhole in Euclidean gravity connecting together separate asymptotic regions. For the moment, such correlation is perfectly reasonable, because the average over couplings explicitly correlates the two systems.

\subsubsection{Computing $\langle z^4\rangle$}\label{sec:z4}
It will also be helpful to understand how to compute $\langle z^4\rangle$. For fixed couplings,
\begin{align}
z^4  = z_{\L}z_{\R}z_{\LL}z_{\RR}= \int \d^{4N}\psi \exp\bigg\{ \i^{q/2}J_{i_1\dots i_q}\big(\psi_{i_1\dots i_q}^\L + \psi_{i_1\dots i_q}^\R+\psi_{i_1\dots i_q}^{\LL}+\psi_{i_1\dots i_q}^{\RR}\big)\bigg\}.
\end{align}
Averaging over $J$, we get
\begin{align}\label{z444}
\langle z^4\rangle = \int \d^{4N}\psi \exp\bigg\{\frac{N}{q}\Big(\frac{\psi_i^a\psi_i^b}{N}\Big)^q\bigg\}.
\end{align}
Here and below, lower case $a,b$ are implicitly summed over the distinct unordered pairs:
\be
(a,b) \in \{(\L,\R), (\LL,\RR), (\L,\LL), (\R,\RR), (\L,\RR),(\R,\LL)\}.
\ee
Eq.~(\ref{z444}) can be given a collective field representation using antisymmetric matrices $G_{ab}$ and $\Sigma_{ab}$, representing correlation between the four replicas: 
\be\label{firstLINE}
\langle z^4\rangle = \frac{\int_{\mathbb{R}}\d G_{ab}\int_{\i\mathbb{R}}\d \Sigma_{ab}}{(2\pi \i/N)^6} \big(\Sigma_{LR}\Sigma_{\LL\RR} - \Sigma_{\L\LL}\Sigma_{\R\RR} + \Sigma_{\L\RR}{\Sigma_{\R\LL}}\big)^N\exp\bigg\{N(-\Sigma_{ab}G_{ab} + \frac{1}{q}G_{ab}^q)\bigg\}.
\ee
Here, the key step of integrating out the fermions was done using the integral
\be
\int \d\psi^1\d\psi^2\d\psi^3\d\psi^4 \exp\Big\{\sum_{1\le a < b \le 4}m_{ab}\psi^a\psi^b \Big\} = \text{Pf}(m) = m_{12}m_{34} - m_{13}m_{24} + m_{14}m_{23}.
\ee
Imitating the steps that led to (\ref{Deriv1}), one can write the result in a way that makes it manifest that the answer only depends on a neighborhood of $G_{ab} = 0$:
\begin{align}\label{z4deriv}
\langle z^4\rangle &= N^{-2N}\left(\partial_{G_{\L\R}}\partial_{G_{\LL\RR}} - \partial_{G_{\L\LL}}\partial_{G_{\R\RR}} + \partial_{G_{\L\RR}}\partial_{G_{\R\LL}}\right)^N \exp\bigg\{\frac{N}{q}G_{ab}^q\bigg\}\Big|_{G_{ab} = 0}\\
&= \frac{N!}{N^{2N}}\left(\frac{N}{q}\right)^{\frac{2N}{q}}\sum_{n_1 +n_2+n_3 = N/q, \  n_i \ge 0} \frac{(qn_1)! (q n_2)! (q n_3)!}{(n_1!)^2 (n_2!)^2 (n_3!)^2}.\label{exactderiv}
\end{align}
In the final expression, we assumed that $N$ is a multiple of $q$. Otherwise, the answer would be zero. When $N$ is large, (and $q > 2$), this sum is dominated by terms where one of the $n_i$ variables is equal to $N/q$, and the others vanish. This gives the answer $\langle z^4\rangle \approx 3\langle z^2\rangle^2$, which is consistent with Gaussian statistics for the $z$ variable at large $N$.

Let's now see how to reproduce this using saddle points. Rotating the contour in (\ref{firstLINE}), and dropping a phase factor that is one if $N$ is a multiple of $q$, one finds the representation
\begin{align}\label{intrepz4}
\langle z^4\rangle &= \int_{\mathbb{R}} \frac{\d^6\sigma_{ab}\d^6 g_{ab}}{(2\pi/N)^6}\exp\bigg\{N\bigg[ \log(\sigma_{\L\R}\sigma_{\LL\RR}  - \sigma_{\L\LL}\sigma_{\R\RR}+\sigma_{\L\RR}\sigma_{\R\LL}) - \i\sigma_{ab}g_{ab} - \frac{1}{q} g_{ab}^q\bigg]\bigg\}.
\end{align}
There are $3\times q\times q$ particularly simple solutions to the saddle point equations that represent ``Wick contractions'' or ``wormhole pairings'' between the four systems. Concretely, we choose one of the patterns of correlation shown here, setting all other $g_{ab}$ components to zero:
\be\label{fig1}
\includegraphics[valign = c, width = .8\textwidth]{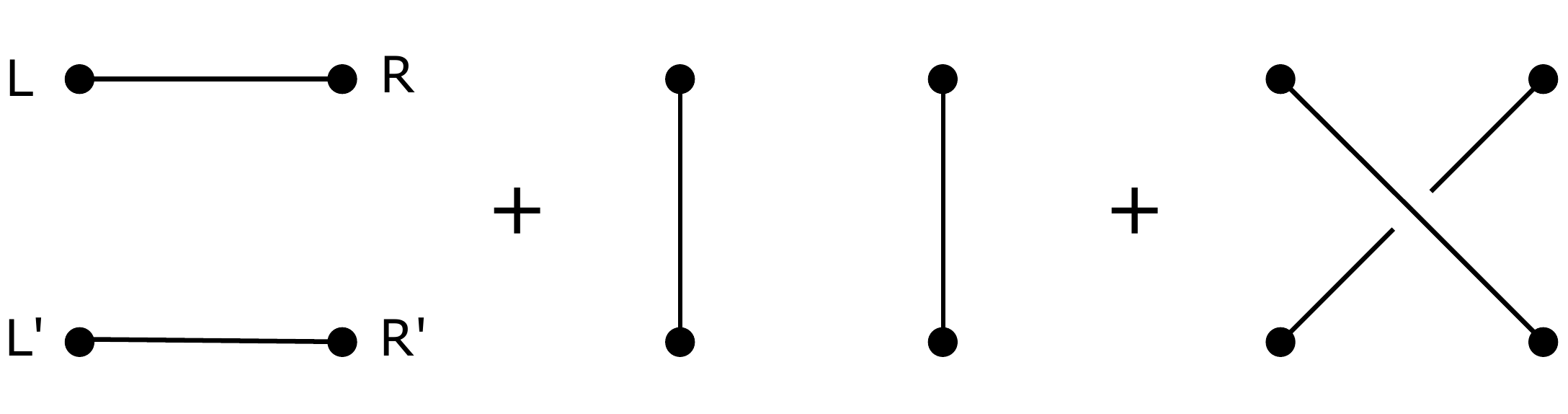}
\ee
For each of the paired systems, one finds the same $q$ saddle points that appeared in the computation of $\langle z^2\rangle$. The three terms in (\ref{fig1}) then lead to the answer $\langle z^4\rangle \approx 3 \langle z^2\rangle$.\footnote{The integral (\ref{intrepz4}) also has saddles where more of the variables are simultaneously nonzero. Among these are some that give subleading contributions, suppressed by $(1/2)^{(q-2)N/q}$ and $(1/3)^{(q-2)N/q}$. These apparently reflect terms in the sum (\ref{exactderiv}) where the total of $N/q$ is split evenly between two or all three of the $n_i$ variables. One also finds a class of saddle points with all $g_{ab}$ nonzero and with an action that competes with the $3q^2$ double wormholes. These saddles would spoil leading-order agreement with (\ref{exactderiv}) and therefore must not contribute. Alternatively, this can be shown using using a generalization of the analysis in appendix \ref{Psiapp}.

 If  saddles like (\ref{fig1}) were the only leading contributions to  the general moments $\langle z^k \rangle$ for arbitrary $k$ then $z$ would be a Gaussian random variable, and the lines would precisely correspond to Wick contractions.  At large $N$ we expect this to be the case (for $q>2$) but have not proven it.}

\subsection{Fixed couplings}\label{subsectionnonaveragedtheory}

 We now turn to our main interest -- studying such systems with {\it fixed} values of the couplings. We will shortly construct a collective field representation for $z^2$ with fixed couplings. The wormhole saddles for $\langle z^2 \rangle$ persist, with weak dependence on the specific $J$ couplings. So the large sample-to-sample fluctuations in $z^2$ must come from other parts of the collective field integral.  We will find that these new contributions are localized at another saddle, which we interpret as a simple example of ``half-wormholes.'' The contribution of this saddle has strong dependence on the specific $J$ matrix, and the sum of the wormhole and the half-wormholes accurately describes  $z^2$.

In existing studies of the SYK model, the $G,\Sigma$ collective fields are introduced after averaging over the couplings, as in the derivation above. However, nothing stops us from introducing them in the theory with fixed couplings. To start, one can write\footnote{We could replace $G^q-(\frac{1}{N}\psi_i^L \psi_i^R)^q$ with $f(G) - f( \frac{1}{N}\psi_i^L \psi_i^R)$ for any function $f$, and the integral over $G$ would still equal one. We have chosen $f$ so that averaging this integral over the $J$ couplings reproduces (\ref{zsquaredintrep1}).}
\be\label{introducingGLR}
z^2 = \int \hspace{-3pt}\d^{2N}\hspace{-3pt}\psi \exp\bigg\{\i^{q/2}\hspace{-18pt}\sum_{1\le i_1<\dots<i_1\le N}\hspace{-20pt}J_{i_1\dots i_q}\left(\psi_{i_1\dots i_q}^\L {+} \psi_{i_1\dots i_q}^\R\right)\hspace{-3pt}\bigg\}\underbrace{\int_{\mathbb{R}}\hspace{-3pt} \d G  \ \delta\big(G {-} \tfrac{1}{N}\psi_i^\L\psi_i^\R\big)\exp\left\{\frac{N}{q}\left[G^q {-} \left(\tfrac{1}{N}\psi_i^\L\psi_i^\R\right)^q\right]\right\}}_{\text{equals 1}}
\ee
We then represent the delta function as a $\Sigma$ integral as in (\ref{introduceDelta}), rotate the contour of $G$ and $\Sigma$ as in (\ref{rotContour}) with $\phi = \frac{\pi}{q}$, and finally interchange the order of integration so that we integrate over the $\sigma$ variable last. It will be convenient to write the answer as
\be\label{integrand}
z^2 = \int_{\mathbb{R}} \d\sigma\, \Psi(\sigma)\Phi(\sigma),
\ee
where the integrand in the final $\sigma$ integral has been split into two factors. The first factor is 
\begin{align}
\Psi(\sigma) &= \int_{\mathbb{R}}\frac{\d g}{(2\pi/N)}\exp\bigg\{N\Big(-\i\sigma g - \frac{1}{q} g^q\Big)\bigg\}.
\end{align}
For the special case $q = 2$, this function is a Gaussian in $\sigma$. For $q = 4,6,\dots$ there is no simple closed form expression, but it can be analyzed by saddle point for large $N$. The function decays faster than exponentially along the real $\sigma$ axis, but slower than a Gaussian, see Appendix \ref{Psiapp}.

The second factor is the more interesting one, since it encodes the dependence on the couplings
\begin{align}\label{intterms}
\Phi(\sigma) &= \int \d^{2N}\psi \exp\bigg\{\i e^{-\frac{\i\pi}{q}}\sigma\psi_i^\L\psi_i^\R + \i^{q/2}J_{i_1\dots i_q}\big(\psi_{i_1\dots i_q}^\L + \psi_{i_1\dots i_q}^\R\big)- \frac{N}{q}\Big(\tfrac{1}{N} \psi_i^\L\psi_i^\R\Big)^q\bigg\}.
\end{align}
This is where the complexity of the fixed-coupling theory is hiding. If we were interested in the average over couplings, we would at this point integrate over $J$ and observe a nice simplification: the interacting terms would cancel and we would be left with
\be\label{averagedPhi}
\langle \Phi(\sigma)\rangle = \int \d^{2N}\psi \exp\bigg\{\i e^{-\frac{\i\pi}{q}}\sigma\psi_i^\L\psi_i^\R\bigg\} = (\i e^{-\frac{\i\pi}{q}}\sigma)^N.
\ee
Substituting this into (\ref{integrand}), one would end up with our old formula for $\langle z^2\rangle$ in (\ref{zsquaredintrep2}).

But we are {\it not} interested in averaging over the couplings. Instead, we would like to understand the theory with a complicated $\Phi(\sigma)$ function that results from a fixed set of couplings. 

\subsubsection{Wormholes persist}
A first question to ask is whether $\Phi(\sigma)$ and $\langle \Phi(\sigma)\rangle$ are actually different. More precisely, in what region of the $\sigma$ plane is $\Phi(\sigma)$ self-averaging? To assess this, one can compare $\langle \Phi(\sigma)\rangle^2$ and $\langle \Phi(\sigma)^2\rangle$. If the two are approximately equal for a given value of $\sigma$, then $\Phi(\sigma)$ is approximately self-averaging.\footnote{A more direct approach would be to choose a set of couplings and just plot $\Phi(\sigma)$. This works for $q = 2$ (see Figure \ref{numphi}) but for $q >2$ it appears to be computationally impractical.} The basic situation is that for sufficiently large $|\sigma|$, the two are equal, but for small $|\sigma|$, they are different.  Crucially, the region where $\Phi(\sigma)$ is self-averaging includes the wormhole saddle points.

To work this out in detail, let's compute $\langle \Phi(\sigma)^2\rangle$. After averaging over $J$, one finds
\begin{align}
\langle \Phi^2(\sigma)\rangle = \int \hspace{-2pt}\d^{4N}\hspace{-2pt}\psi \exp\bigg\{\i e^{-\frac{\i\pi}{q}}\sigma(\psi_i^\L\psi_i^\R + \psi_i^{\LL}\psi_i^{\RR}) + \frac{N}{q}\bigg[\Big(\frac{\psi_i^\L\psi_i^{\LL}}{N}\hspace{-2pt}\Big)^{\hspace{-2pt}q}{+}\Big(\frac{\psi_i^\R\psi_i^{\RR}}{N}\hspace{-2pt}\Big)^{\hspace{-2pt}q}{+}\Big(\frac{\psi_i^\L\psi_i^{\RR}}{N}\hspace{-2pt}\Big)^{\hspace{-2pt}q}{+}\Big(\frac{\psi_i^\R\psi_i^{\LL}}{N}\Big)^q\bigg]\bigg\}\notag
\end{align}
The collective-field representation of this is similar to the one for $\langle z^4\rangle$, except that we freeze $\sigma_{\L\R} = \sigma_{\LL\RR} = \sigma$ and set $g_{\L\R} = g_{\LL\RR} = 0$. The analog of (\ref{exactderiv}) is the exact formula
\begin{align}\label{exactphi2}
\langle \Phi(\sigma)^2\rangle = \sum_{n_1+n_2+n_3 = \frac{N}{q}, \ n_i\ge 0} \frac{N!}{N^{2q(n_2+n_3)}}\left(\frac{N}{q}\right)^{2(n_2+n_3)}\frac{\sigma^{2qn_1}}{(qn_1)!}\frac{(q n_2)!(q n_3)!}{(n_2!)^2(n_3!)^2}.
\end{align}
This function is easy to plot numerically for reasonably large values of $N$, and by comparing to (\ref{averagedPhi}) one can determine the region in which $\Phi(\sigma)$ is self-averaging.

One can also figure out the answer directly in the large $N$ limit using saddle points of the collective field description. The analog of (\ref{intrepz4}) for this case is
\begin{align}\label{Phisquared}
\langle \Phi^2(\sigma)\rangle &= \hspace{-3pt}\int_{\mathbb{R}} \hspace{-3pt}\frac{\d^4\sigma_{AB}\d^4 g_{AB}}{(2\pi/N)^4}\exp\bigg\{\hspace{-3pt}N\bigg[\log(\sigma^2 {-} \sigma_{\L\LL}\sigma_{\R\RR}{+} \sigma_{\L\RR}\sigma_{\R\LL} ) - \i\sigma_{AB}g_{AB} - \frac{1}{q} g_{AB}^q\bigg]\hspace{-2pt}\bigg\}.
\end{align}
Here, we are using a convention that the $A,B$ index pair is summed over the four values
\be
(A,B) \in \{(\L,\LL), (\R,\RR), (\L,\RR),(\R,\LL)\}.
\ee
For any value of $\sigma$, this integral always has a trivial saddle point at $\sigma_{AB} = g_{AB} = 0$. This saddle point contributes $\langle \Phi(\sigma)^2\rangle \supset \langle \Phi(\sigma)\rangle^2$. So {\it if} this trivial saddle dominates, we can conclude that $\Phi(\sigma)$ is self-averaging.

First consider the case $\sigma = 0$. Then the contribution of the trivial saddle point is zero, and $\Phi(\sigma)$ is definitely {\it not} self-averaging. Instead, the leading contribution to the integral comes from $2\times q\times q$ saddle points corresponding to the second and third patterns of correlation in Figure (\ref{fig1}). The factor of two in this counting corresponds to the choice between these two patterns, and to fix notation we will focus on the third pattern, so that that $\sigma_{\L\RR}$ and $\sigma_{\R\LL}$ are the nonzero ones. The factor of $q\times q$ corresponds to the independent choices of phase for $\sigma_{\L\RR}$ and $\sigma_{\R\LL}$ (and the corresponding $g_{AB}$ variables). It will be helpful to group the saddles into $q$ groups of $q$ each, such that the product of the phases is $\sigma_{\L\RR}\sigma_{\R\LL} = e^{\frac{2\pi \i m}{q}}$.

Now, suppose that we deform $\sigma^2$ away from zero. These saddles will deform to new locations such that $\sigma_{\L\RR}\sigma_{\R\LL} = e^{\frac{2\pi \i m}{q}}s^2$, with $s$ initially close to one. For a fixed value of $s$, we can solve some of the saddle point equations in order to eliminate the $g_{AB}$ variables in terms of $s$. For such a configuration, the action is:\footnote{To do this one has to choose a branch of the solution for the $g$ variables. We have chosen here the branch such that there is a saddle point at $s = 1$ in the case with $\sigma  = 0$.}
\be\label{cont1}
\exp\bigg\{N\Big[\log(\sigma^2 + e^{\frac{2\pi \i m}{q}}s^2) - 2\frac{q-1}{q} s^{\frac{q}{q-1}}\Big]\bigg\}.
\ee
When $\sigma = 0$, saddles with all values of $m$ contribute equally. However, when $\sigma$ is nonzero, the degeneracy is lifted, favoring the saddles where $e^{\frac{2\pi \i m}{q}}$ points in the same direction as $\sigma^2$. For example, if $\sigma^2$ is close to the positive real axis (more precisely, if the phase of $\sigma^2 = e^{i\phi}|\sigma|^2$ satisfies $|\phi| \le \frac{\pi}{q}$), then the $m = 0$ saddles dominate, and the value of $s$ is determined by solving
\be\label{saddleS}
\frac{s}{\sigma^2 + s^2} = s^{\frac{1}{q-1}}
\ee
and choosing the  branch such that $s = 1$ when $\sigma = 0$. For larger values of the phase $\phi$, the calculation is similar, but the dominant family of saddles are the ones with the value of $m$ so that $2\pi m/q$ is closest to $\phi$.

\begin{figure}
\begin{center}
\includegraphics[width = .6\textwidth]{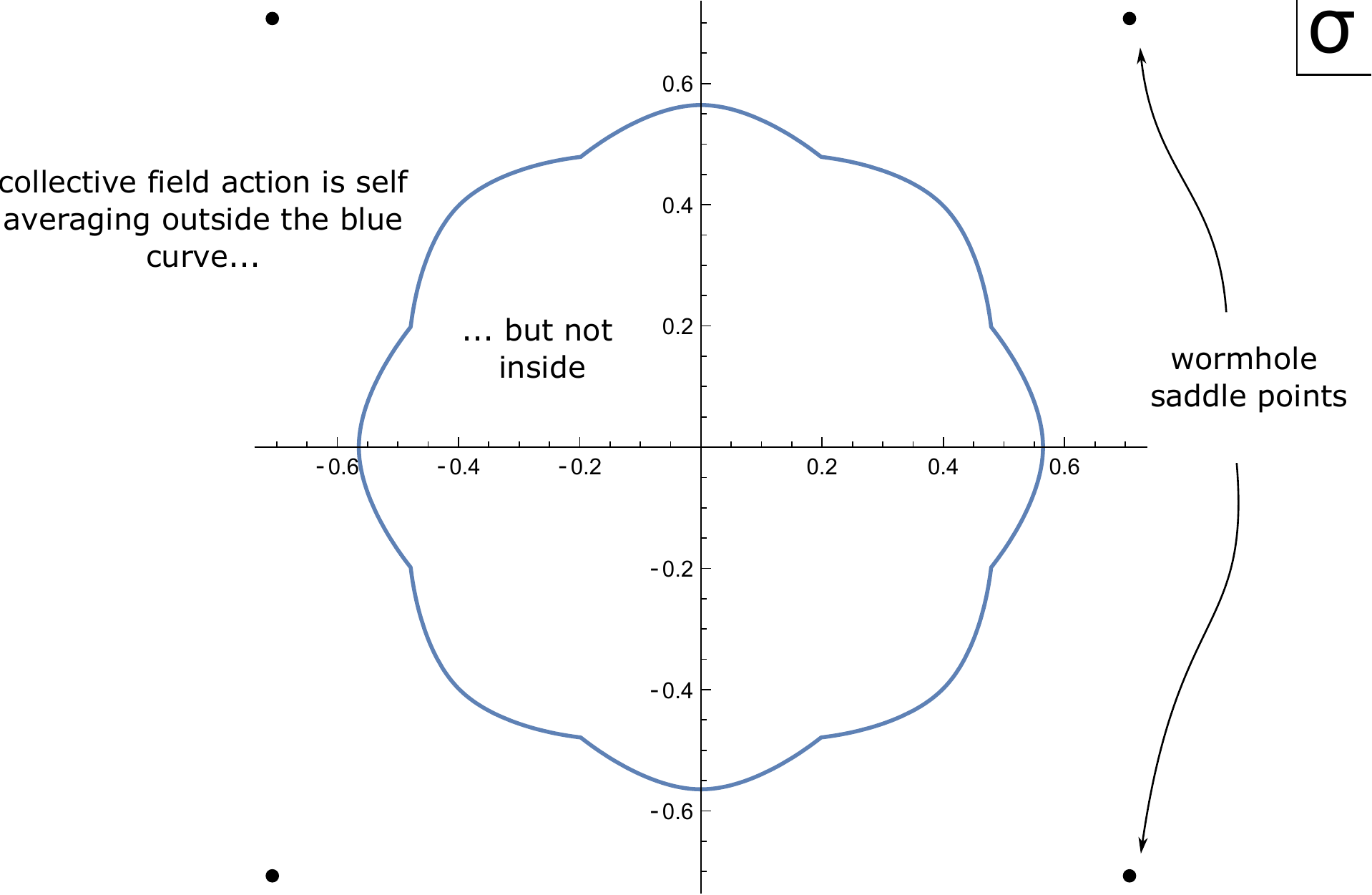}
\caption{{\sf We plot the complex $\sigma$ plane. Outside the blue scalloped curve, the trivial saddle point dominates, and $\Phi(\sigma)$ is self-averaging. The black dots are the locations of the ``wormhole'' saddle points. They are in the self-averaging region: wormholes persist.}}\label{figPolar}
\end{center}
\end{figure}

By solving (\ref{saddleS}), computing the action, and comparing it to the contribution of the trivial saddle, we can determine in what regions $\Phi(\sigma)$ is self-averaging. If the trivial saddle wins, it is self-averaging. If the nontrivial saddle just described wins, it is not. We show a plot in Figure \ref{figPolar} for the case of $q = 4$.

Recall that the ``wormhole'' saddles of $\langle z^2\rangle$ are at values of $\sigma$ on the unit circle.\footnote{Just looking at the figure, it might be hard to imagine why all four wormhole saddles should be included. This is explained in appendix \ref{Psiapp}.} This turns out to be well within the self-averaging region of the $\sigma$ plane, where the nontrivial saddle point is exponentially subdominant. So in fact the leading correction to self-averaging comes from perturbative fluctuations of the $g_{AB}$ and $\sigma_{AB}$ fields around the trivial $g_{AB} = \sigma_{AB} = $ saddle point. One finds 
	\be
	\langle \Phi(\text{WH saddle})\rangle = 1, \hspace{20pt} \langle \Phi^2(\text{WH saddle})\rangle = 1 + \frac{2\cdot q!}{q^2}\frac{1}{N^{q-2}} +\dots
	\ee
So the self-averaging is not perfect, but it is good if $N$ is large and $q > 2$. We conclude that the wormhole saddle points survive intact in the non-averaged theory.

\subsubsection{Half-wormholes exist}
By themselves, these wormholes cannot be an accurate approximation to $z^2$, because they are self-averaging while $z^2$ itself is not. In fact, there is another contribution, roughly equal in magnitude to the wormhole contributions, coming from the non-self-averaging region near the origin of the $\sigma$ plane. In the averaged theory, this region contributes very little to the integral, but for typical fixed values of the couplings, it it important. We will refer to this contribution as coming from ``half-wormholes'' for reasons that will be explained below.

Because the $\Psi(\sigma)$ function rapidly decays away from the origin, one can show that the integral over the non-self-averaging region can be well approximated by just the contribution at $\sigma = 0$. So the semiclassical approximation for fixed couplings looks like
\be\label{Structure}
z^2 \approx \Big(\text{wormhole saddles with $|\sigma| = 1$}\Big) + \Big(\text{``half-wormholes'' saddle at $\sigma = 0$}\Big).
\ee
The first term is self-averaging and the second term depends strongly on the couplings. The approximation can be systematically improved by including fluctuations around these saddles.

For a typical realization, the two contributions are of the same order. In a leading approximation, the half-wormholes contribution is just $\Phi(0)$, because the $\Psi$ function acts approximately as a delta function (see (\ref{normapp})). The RMS value of the half-wormholes contribution is then determined by (\ref{cont1}) with $s = 1$ and $\sigma = 0$, or by the exact formula (\ref{exactphi2}). Either way, one finds
	\be\label{sameOrder}
		\sqrt{\langle \Phi^2(0)\rangle} \sim  \langle z^2\rangle
	\ee
	which is approximately equal to the contribution of the wormhole saddles. 

\subsubsection{Together, wormholes plus half-wormholes are a good approximation}\label{error223}

We now give an independent check of this approximation, using a method we will return to later in section \ref{sectiongeneral}. Denote  the RHS of \eqref{Structure} by ${\mathbf{z^2}}$.   Then define the error in this approximation by $\text{Error} = z^2 - \mathbf{z^2}$.   We want to show that $\text{Error}$ is small in the ensemble, so we compute its moments.   First we examine its average:
\be\label{errorone}
\langle \text{Error}\rangle = \langle z^2 \rangle - \langle \mathbf{z^2} \rangle.
\ee
Because $\langle \Phi(0)\rangle = 0$, the average of the $\sigma = 0$ part of $\mathbf{z^2}$ vanishes.   The average of the $| \sigma | = 1$ part of $\mathbf{z^2}$   is the expectation value of the self-averaging wormhole saddles, which gives $\langle z^2 \rangle$ in the semiclassical approximation. So \eqref{errorone} vanishes.

A stronger test is to compute the average of the second moment
\be\label{errortwo}
\langle \text{Error} \cdot \text{Error} \rangle  = \langle z^2 \cdot z^2 \rangle -  \langle z^2 \cdot \mathbf{z^2} \rangle - \langle \mathbf{z^2} \cdot z^2 \rangle + \langle \mathbf{z^2} \cdot \mathbf{z^2} \rangle.
\ee
We can think of the first factor in each of the terms in \eqref{errortwo} as describing the original system, and the second factor as describing a second, auxiliary system included to calculate the second moment.   The first term on the RHS of \eqref{errortwo} is just the fourth moment of $z$ computed in section \ref{sec:z4}.  The saddle points determining this quantity at large $N$ are pictured in (\ref{fig1}).  Here the points $\L,\R$ comprise the original system, the points $\LL, \RR$ are the auxiliary system.  

To calculate $\langle z^2 \cdot z^2 \rangle$ we introduce collective fields $g_{\L\R}, \sigma_{\L\R}$ linking the points in the original system, collective fields $g_{\LL\RR}, \sigma_{\LL\RR}$ linking the points in the auxiliary system, and collective fields $g_{AB}, \sigma_{AB}$ linking the original system to the auxiliary system.  The first pattern in (\ref{fig1}) describes a wormhole in the original system and one in the auxiliary system. The second and third patterns describe wormholes linking the original to the auxiliary system. 

  If we just focus on the original system, which is our interest, the first pattern is a wormhole, but the second and third patterns only contain the remnant of the wormhole connecting to the auxiliary system, described by the collective fields $g_{\L\R}, \sigma_{\L\R}$ assuming the value of that wormhole saddle.  We call such a configuration a ``half-wormhole" and illustrate the situation as follows:
\be\label{fighalfwormholesz}
\includegraphics[valign = c, width = .8\textwidth]{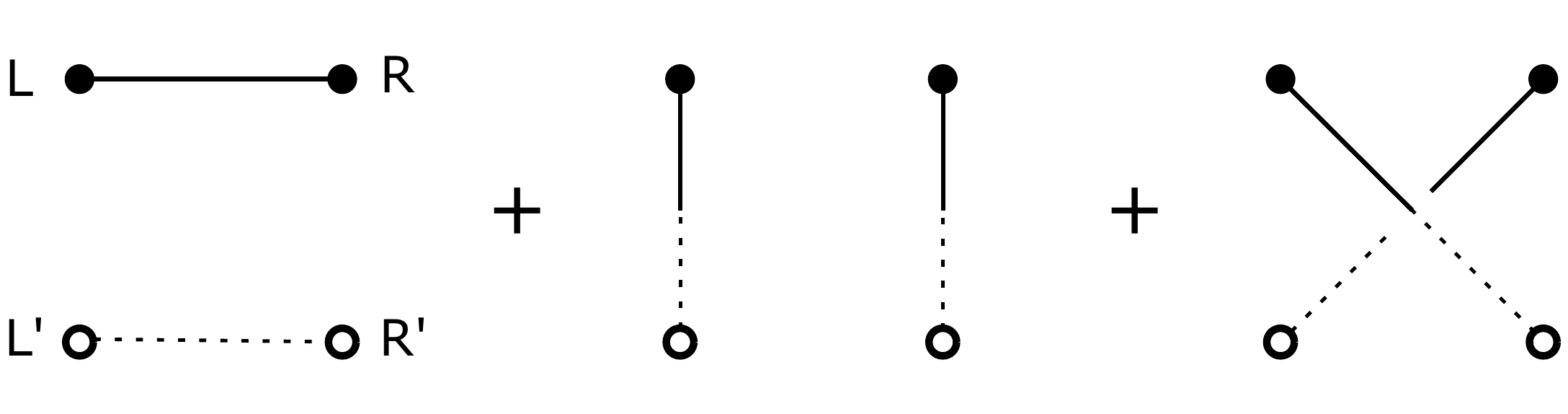}
\ee

We now turn to evaluating the other terms in \eqref{errortwo} semiclassically, starting with   $\mathbf{z^2} \cdot \mathbf{z^2}$.   To understand the variable $\sigma$ used in \eqref{Structure}  in terms of the $\sigma_{ab}$ variables used to calculate $\langle z^4 \rangle$ we combine equations \eqref{intrepz4}, \eqref{integrand},\eqref{intterms} and\eqref{Phisquared} to arrive at 
\be\label{z545}
\langle z^4\rangle = \int \d\sigma_{\L\R}\d\sigma_{\LL\RR}\langle \Phi^2(\sqrt{\sigma_{\L\R}\sigma_{\LL\RR}})\rangle \Psi(\sigma_{\L\R})\Psi(\sigma_{\LL\RR}).
\ee
We see that $\sigma^2$ should be identified with $\sigma_{\L\R}\sigma_{\LL\RR}$.   
The four terms in $\mathbf{z^2} \cdot \mathbf{z^2}$ correspond to doing the collective field integral \eqref{intrepz4} with the $\sigma_{LR}$ and $\sigma_{L'R'}$ variables fixed at their values in the wormhole or half-wormhole saddle points (or integrated in a neighborhood of them). Naively this should not make much difference, since the integral is dominated by the contributions of these saddle points anyhow, so we tentatively would conclude $\mathbf{z^2} \cdot \mathbf{z^2}\approx \langle z^4  \rangle$. In principle, though, something could go wrong at this step, because (i) fixing some variables to their saddle point values could change the saddle point contour analysis for the remaining variables and (ii) there are wormhole-half-wormhole cross terms in $\langle \mathbf{z}^2\cdot \mathbf{z}^2\rangle$ that do not appear in the saddle point analysis of $\langle z^4\rangle$. However, in this simple  case we can confirm by comparison to the exact formula \eqref{exactphi2} that the contour problem does not arise and the cross-terms are small.\footnote{The cross-terms can explicitly be seen to be small by plugging in the wormhole and half-wormhole values for $\sigma_{\L\R}$ and $\sigma_{\LL\RR}$ in (\ref{z545}). The smallness of this contribution is intuitive because the cross-terms represent the correlation between the contributions of the wormhole and the strongly fluctuating half-wormhole. Such correlations are small compared to the diagonal terms.}, so indeed $\mathbf{z^2} \cdot \mathbf{z^2}\approx \langle z^4  \rangle$.

After arguing similarly for the other two terms in \eqref{errortwo}, $\langle \mathbf{z^2} \cdot z^2 \rangle$ and $\langle z^2 \cdot \mathbf{z^2} \rangle$, we conclude that all four terms are approximately equal in magnitude, so $\langle \text{Error}^2\rangle$ is small. (How small depends on how many orders in small fluctuations around the saddle points are included in (\ref{Structure}).) This ensures that the approximation \eqref{Structure} is accurate for any member of the ensemble (a set of $J$ couplings) that occurs with high probability -- a ``typical" member.  Of course there are highly atypical members where the error is large.

\subsubsection{Multiple descriptions coexist}
In the above calculations, we insisted on computing $z^2$ in a formalism with explicit $LR$ collective fields (again, in this model there is no $G_{LL}$ or $G_{RR}$, so we left off the subscripts). This made it logically possible for a wormhole to contribute, and indeed we found that it does. 

But of course this is not necessary. Another possibility is to compute $z^2$ by doing two separate computations of $z$, with no $LR$ collective fields. So there are two ways of  computing $z^2$. If we carry out both exactly, then they are exactly equal. But in a semiclassical approximation, the equivalence is approximate and nontrivial. For example, in the leading semiclassical approximation, the statement of equivalence is
\be
z^2 \approx \langle z^2\rangle + \Phi(0).
\ee
Writing both sides out in terms of fermions, this is
\begin{align}\label{identity}
\int \d^{2N}\psi \exp\bigg\{ \i^{q/2}J_{i_1\dots i_q}\big(\psi_{i_1\dots i_q}^\L + \psi_{i_1\dots i_q}^\R\big)\bigg\} \approx &\int \d^{2N}\psi\exp\bigg\{\frac{N}{q}\Big(\tfrac{1}{N}\sum_{i = 1}^N \psi_i^\L\psi_i^\R\Big)^q\bigg\}  \\
&\hspace{-50pt}+ \int \d^{2N}\psi \exp\bigg\{ \i^{q/2}J_{i_1\dots i_q}\big(\psi_{i_1\dots i_q}^\L + \psi_{i_1\dots i_q}^\R\big)- \frac{N}{q}\Big(\tfrac{1}{N} \psi_i^\L\psi_i^\R\Big)^q\bigg\}.\notag
\end{align}
As far as we are aware, the most efficient way to derive it is actually the series of steps we followed above. To recap, the steps are as follows: (i) introduce the $G = G_{LR}$ variable by inserting 
\be
1 = \int_{\mathbb{R}} \d G  \ \delta\big(G - \tfrac{1}{N}\psi_i^\L\psi_i^\R\big)\exp\left\{\frac{N}{q}\left[G^q - \left(\tfrac{1}{N}\psi_i^\L\psi_i^\R\right)^q\right]\right\}
\ee
inside the integral on the LHS. (ii) represent the delta function using a $\Sigma$ variable. (iii) rotate the contour and approximate the resulting $G,\Sigma$ integral by saddle points. (iv) further approximate the wormhole saddle points by their average value.\footnote{This step could be omitted at the cost of a slightly more complicated approximate equality.}
 
At large $N$, one can compute $\langle (LHS -RHS)^2\rangle / \langle LHS^2\rangle = \frac{4}{3}\frac{q!}{q^2}N^{2-q}$, indicating that the error is small but not exponentially so. This error can be systematically reduced by undoing step (iv) above and including further terms in the semiclassical approximation around the wormhole and half-wormhole saddles.

For some purposes, the LHS may be a simpler description than the RHS. But there are cases where the RHS can be useful, mainly because the first term (the wormhole) is simple. For example, if we are interested in the average over the couplings, the final term (the half-wormholes) contribute zero, so the RHS reduces right away to the wormhole. Second, although the two terms on the RHS are typically of the same order, there are interesting ways of modifying the calculation so that the wormhole piece dominates. We will give an example of that next.

\subsection{Adding a coupling between the replicas}
We can shift the balance between self-averaging and non-self-averaging contributions to (\ref{Structure}) by considering a generalization where the two replicas are coupled together:
\be\label{defzeta}
\zeta(\mu) = \int \d^{2N}\psi\exp\bigg\{\mu\sum_{i = 1}^N\psi_i^\L\psi_i^\R + \i^{q/2}\hspace{-15pt}\sum_{1\le i_1<\dots<i_1\le N}J_{i_1\dots i_q}\left(\psi_{i_q\dots i_q}^\L + \psi_{i_1\dots i_q}^\R\right)\bigg\}.
\ee
Note that when the coupling $\mu$ is zero, this is just the square of $z$:
\be
\zeta(0) = z^2.
\ee
One can think of the system with nonzero $\mu$ as being somewhat analogous to the Maldacena-Qi eternal traversable wormhole \cite{Maldacena:2018lmt}. One can also think of it as being similar to a ``two instant of time'' version of a single copy of SYK.

Either way, one can write a formula for the non-averaged $\zeta(\mu)$ quantity using the same $\Psi$ and $\Phi$ functions that appear in the computation of $z^2$:
\be\label{newIntegrand}
\zeta(\mu) = \int_{\mathbb{R}}\d\sigma \Psi(\sigma)\Phi(\sigma - \i e^{\frac{\i\pi}{q}}\mu).
\ee
We would like to understand what happens to the structure (\ref{Structure}) when $\mu$ is nonzero. It is particularly easy to do this when $\mu$ is small. In that case, one can use first-order perturbation theory, just evaluating the new integrand (\ref{newIntegrand}) on the ``wormhole'' and ``half-wormhole'' saddle points. The $\mu$ perturbation breaks the degeneracy between the saddles with different phases. Keeping only the saddles that dominate for positive real $\mu$, one finds (for $\mu \ll 1$)
\be\label{zetaGuess}
\zeta(\mu) = \Big(\text{self-averaging saddle with $|\sigma| = 1$}\Big)e^{N\mu} + \Big(\text{fluctuating piece from $\sigma = 0$}\Big) e^{N \mu^2}.
\ee
In this expression, the quantities in parentheses are in the $\mu = 0$ theory, so in particular, the two pieces are typically of the same order. Once we include the $\mu$ perturbation, the self-averaging saddle point is enhanced relative to the fluctuating piece. So the mechanism by which $\zeta(\mu)$ becomes self-averaging is simply that the self-averaging part of the integration space is enhanced relative to the fluctuating part.\footnote{We expect that a similar mechanism will explain the self-averaging property of the replica wormhole saddle in the SYK version of that story \cite{Penington:2019kki} with fixed fermion couplings.  The shared radiation bath serves to couple the various replicas together.}

We avoided labeling the first term in (\ref{zetaGuess}) as a wormhole because whether we think is as a ``wormhole'' or a ``disk'' depends in this case on whether we think of $\zeta(\mu)$ as being analogous to two coupled copies of SYK, or a single SYK theory with two instants of time. 

In the next section we will discuss the full SYK model, and to make the transition to that more complicated case, it is helpful to think of $\zeta(\mu)$ as being analogous to a single SYK partition function. Then the two contributions to $\zeta(\mu)$  in (\ref{zetaGuess}) are analogous to the disk and half-wormhole contributions. In a computation of $\zeta(\mu)^2$ for fixed couplings, we would have $(\text{disk})^2$, $(\text{half-wormhole})^2$, $(\text{disk})\times (\text{half-wormhole})$ and wormhole contributions.

\section{Extrapolation to regular SYK}\label{sectiongeneral}

In the previous section, we studied the quantity $z^2$ for a fixed set of couplings, using a path integral over $G,\Sigma$ collective fields. $z^2$ of course factorizes into a product of the two $z$'s, but this is not obvious from the point of view of the semiclassical evaluation (\ref{Structure}). We found that the ``wormhole'' saddle point, which leads to a non-factorized answer for the average $\langle z^2\rangle$, survives in the non-averaged theory. Though the wormhole threatens to spoil factorization, it is restored by an additional contribution from ``half-wormholes,'' which live in the non-self-averaging region of the integrand. 

Our goal for this section is to address this problem in a more general setting, and to study how the semiclassical evaluation of the $G,\Sigma$ integral for a product $Z_L Z_R$ factorizes. We can write such integrals even without averaging over the couplings, by generalizing the steps in section \ref{subsectionnonaveragedtheory}. 

In fact there are two natural choices for a $G,\Sigma$ integral description of the product $Z_L Z_R$; one representation includes $G_{LR}$ collective fields and does not manifestly factorize, the other does not include the $LR$ collective fields, and is manifestly equal to the product of $G,\Sigma$ integral representations for the individual factors. As a result, we find different semiclassical expressions for the product $Z_L Z_R$, evaluated with or without using the $LR$ collective fields. Each expression includes contributions from half-wormholes, but only one includes wormhole contributions from the $LR$ collective fields.

The factorization problem is now twofold: first one must understand how to evaluate both the product $Z_L Z_R$ and the individual factors semiclassically, including both self-averaging and non-self-averaging contributions. Then one must understand how the wormhole and half-wormhole contributions in $Z_L Z_R$ conspire to factorize into the product of expressions for the $Z_L$ and $Z_R$. 

To state this problem more precisely, we introduce a notation which fixes this ambiguity of the $G,\Sigma$ integral representation. We define the quantity $\Z(Z)$ to be a specific choice of $G,\Sigma$ integral representation for its argument $Z$. Here $Z$ is a placeholder symbol which represents a partition function-like quantity, such as a partition function or a product of partition functions. We often suppress the argument when it is unimportant or clear from context. In particular, we choose to associate $\Z(Z)$ with the $G,\Sigma$ integral for $Z$ with all of the collective field variables introduced. So, for example, when $Z$ is a product $Z_L Z_R$ of partition functions we have 
\be\label{gsigmanotfactorized}
\Z(Z_L Z_R) \equiv \int \D G_{LL} \D G_{RR} \D G_{LR}\D \Sigma_{LL} \D \Sigma_{RR} \D \Sigma_{LR} \; e^{- I_{fl}^{(L,R)} }.
\ee
Note that this integral is defined to include an integral over the $LR$ collective fields as well as $LL$ and $RR$. Here the action $I_{fl}^{(L,R)}$ is the $G,\Sigma$ action for a fixed choice of couplings, including a generalization of the function $\Phi(\sigma)$ discussed in the previous section.

This expression should be contrasted with 
\be\label{gsigmafactorized}
\Z(Z_L) \Z(Z_R) \equiv \bigg( \int \D G_{LL} \D \Sigma_{LL} \; e^{- I^{(L)}_{fl}} \bigg)\bigg( \int \D G_{RR} \D \Sigma_{RR} \; e^{- I^{(R)}_{fl}} \bigg).
\ee
The factorization problem in SYK can then be expressed as an equality between these two integrals,
\be\label{factorizationequation}
\Z(Z_L Z_R) = \Z(Z_L) \Z(Z_R).
\ee
These two representations of $Z_L Z_R$ are somewhat trivially related; we can go from the first expression to the second by integrating out the $LR$ variables along their defining contour. This essentially undoes the procedure used to introduce the $LR$ variables in the first place. Likewise, the $LR$ variables can be trivially reintroduced in the second expression. The more interesting question is to ask how this factorization property can be understood \textit{semiclassically}. What are the saddle points in (\ref{gsigmanotfactorized}) and (\ref{gsigmafactorized}), and how do they conspire to approximately reproduce (\ref{factorizationequation})? 

\subsection{Proposal}
Our proposal is that to evaluate a general $\Z(Z) = \Z$ with fixed couplings, one has to sum over configurations that appear as ``half'' of saddle points in an averaged computation of the square of this quantity, $\langle \Z^2\rangle = \langle \Z_1\Z_2\rangle$.\footnote{Here and below, we assume $\Z$ is real. If $\Z$ is complex we should study $|\Z|^2$ instead.} So, for example, if we were interested in evaluating $\Z(Z_L)$ for fixed couplings, we would consider the computation of $\langle \Z(Z_L)_1\Z(Z_L)_2\rangle$. And if we were interested in computing $\Z(Z_LZ_R)$, we would consider $\langle \Z(Z_LZ_R)_1\Z(Z_LZ_R)_2\rangle$. In either case, the $G,\Sigma$ fields describing the average of the square, $\langle \Z_1\Z_2\rangle$, have 11, 22, and 12 components. 

We propose that the saddle points contributing to $\Z$ with fixed couplings are the 11 (or 22) components of saddle points for $\langle \Z_1\Z_2\rangle$. There are two subcases.

\begin{enumerate}
	\item If we start with a saddle point for $\langle \Z_1\Z_2\rangle$ with vanishing 12 correlation, it leads to a 11 configuration that is a saddle point of $\langle \Z\rangle$. Such a configuration is indeed an obvious guess for a semiclassical evaluation of $\Z$. Although we do not know how general it is that such saddles  survive in the fixed-coupling theory, we have seen in section \ref{sectiononepoint} that even wormhole saddle points can do so.

\item If we start with a saddle point with nonzero 12 correlation, then the resulting 11 configuration will not be a saddle point of $\langle \Z\rangle$, but it can be a non-self-averaging saddle point for $\Z$ with fixed couplings. These configurations are not such obvious guesses, but they play a crucial role. We refer to them as half-wormholes, and they generalize the $\sigma = 0$ contribution in the model from section \ref{sectiononepoint}.
\end{enumerate}

\subsection{Consistency check}
The proposal described above is motivated by the example from section \ref{sectiononepoint}, but we can also sketch a general plausibility argument for it. As in section \ref{error223}, the idea is to define an ``Error'' as the difference between our proposed semiclassical answer and the exact answer, and to check that $\langle \text{Error}^2\rangle$ is small.

Let $Z$ be a partition function or product of partition functions in an SYK-like theory, with a collective field representation
\be\label{zintegralgeneral}
\Z = \Z(Z) = \int \D x \;e^{-I_{fl}(x)}.
\ee
The weight $I_{fl}(x)$ is the action for a fixed set of random couplings, and the integration variable $x$ may include $G,\Sigma$ variables for a single replica, or a matrix of $G,\Sigma$ variables for multiple replicas.\footnote{For $z^2$ the analog is $e^{-I_{fl}(g,\sigma)} = \Phi(\sigma)e^{N(-\i g \sigma - \frac{1}{q}g^q)}$.}

The averaged square of $\Z$, $\langle \Z^2\rangle = \langle \Z_1 \Z_2\rangle$, is also computed by an integral over collective field variables. In this case, these variables consist of $x_1$ and $x_2$, corresponding to copies of the variables used in (\ref{zintegralgeneral}), as well as variables $x_{12}$. We use $X=(x_1, x_2, x_{12})$ to denote the set of all of these variables.

$\langle \Z_1 \Z_2 \rangle$ can be expressed as an integral over $X$ with the appropriate action $I_{ave}(X)$,
\be\label{z2aveintegralgeneral}
\langle \Z_1 \Z_2 \rangle = \int \D X\; e^{-I_{ave}(X)}.
\ee
This integral can be approximated by a sum over saddle points, $X=X_I^* \equiv (x_{1,I}^*, x_{2,I}^*, x_{12,I}^*)$:
\be\label{twozsaddlessum}
\langle \Z_1 \Z_2 \rangle  \approx \sum_I e^{-I_{ave}(X^*_I)}.
\ee
Here and below, we are suppressing one-loop factors, although they can be included.

To implement the proposal above, we now try to approximate $\Z$ by a sum \label{zdefthree} $\Zs$ over configurations which correspond to the saddle points $X_I^*$. More precisely, we sum over the $x_1$ components of those saddle points:\footnote{The prime on the sum means that we sum over distinct $x_1$ components of the saddles. This is important because there may be multiple saddle points for which the $x_1$ components are the same, but the $x_2$ and/or $x_{12}$ components are different. For example, the double cone saddle point in SYK has a family of saddle points, labeled by $\delta t$. In this family, the $x_1$ and $x_2$ components are independent of $\delta t$.}
\begin{align}\label{Zansatz}
\Zs &\approx  \sum_I{}^{'} e^{-I_{fl}(x = x_{1,I}^*)} = \sum_I{}^{'} \int \D x \; e^{-I_{fl}(x)} \; \delta(x- x_{1,I}^*).
\end{align}
Now, define $\text{Error} \equiv  \Z-\Zs$ as the error in this approximation, and study $\langle \text{Error}^2\rangle = \langle \text{Error}_1\; \text{Error}_2\rangle$:
\be\label{erroreq}
\langle \text{Error}_1 \; \text{Error}_2\rangle = \langle \Z_1 \Z_2\rangle  - 2\langle \Z_1 \Zs_2 \rangle+ \langle \Zs_1 \Zs_2 \rangle.
\ee
Using equations (\ref{zintegralgeneral}) and (\ref{Zansatz}), and simplifying terms using
\be
 \langle e^{-I_{fl}(x_1) - I_{fl}(x_2)}\rangle = \int \D x_{12} \;e^{-I_{ave}(X)},
\ee
which follows from (\ref{z2aveintegralgeneral}), the three terms in (\ref{erroreq}) can be written as
\begin{align}\label{threetermssecond}
\langle \Z_1 \Z_2\rangle & = \int \D X \; e^{-I_{ave}(X)}
\cr
\langle \Z_1 \Zs_2 \rangle & =\sum_{I}{}^{'} \int \D X \; e^{-I_{ave}(X)}\; \delta(x_2 - x_{2,I}^*)
\cr
\langle \Zs_1 \Zs_2 \rangle & =\sum_{I,J}{}^{'} \int \D X \; e^{-I_{ave}(X)}\; \delta(x_1 - x_{1,I}^*)\delta(x_2 - x_{2,J}^*).
\end{align}
The only difference between these expressions is that in some cases, the $x_1$ and $x_2$ variables are fixed at saddle point values,\footnote{In order for the Error to be small, we need to include the integral over fluctuations around the saddle points, to an order that depends on tolerance. The argument can and should be extended to include such fluctuations, but the notation is awkward so we avoided it here.} instead of freely integrated over. It therefore seems plausible that the three terms are approximately equal, and therefore (\ref{erroreq}) is small. 

However, this is not a proof, for two reasons. First, as a general fact about multidimensional integrals, restricting some integration variables to their values at contributing saddle points can sometimes lead to a bad approximation to an integral, due to the possibility that the contour/thimble analysis for the remaining variables can be affected by fixing some of the variables. Second, $\langle \Zs_1 \Zs_2 \rangle$ will contain cross-terms in which the $x_1$ and $x_2$ variables are fixed to different saddle points, and for the argument to work these contributions need to be subleading. In the $z^2$ model we saw explicitly that these issues do not arise. We don't have a general justification, but  the above argument does show that if indeed the same saddles dominate in all three terms, the approximation is good (or at least can be made good by including more terms in a systematic expansion about the saddles).

\subsection{Application to factorization}

We can apply this proposal to the factorization problem by comparing the semiclassical formulas for $\Z(Z_L Z_R )$ and $\Z(Z_L )\Z(Z_R )$. The formula for $\Z(Z_L Z_R )$ includes contributions from wormholes connecting $Z_L$ and $Z_R$, as well as from half-wormholes which correspond to “half” of wormhole saddle points in $\langle \Z(Z_L Z_R)_1 \Z(Z_L Z_R)_2\rangle$. The formula for $\Z(Z_L )\Z(Z_R )$ includes related half-wormholes, but no complete wormhole. The equality of the two computations, and therefore factorization, implies a nontrivial relationship between the contributions of wormholes and half-wormholes.

In this section we illustrate this using an example from the SYK model. Our choice of $Z$ for this example is $Y(T)$, defined in \cite{Saad:2018bqo}. $Y(T)$ can be thought of as a microcanonical version of the analytically continued partition function $Z(\beta+ i T)$.

Averages of products of $Y(T)$ and its complex conjugate receive contributions from the ``double cone'' wormhole saddle point.\footnote{For sufficiently long times $T$.} For example, $\langle Y(T) Y(T)^*\rangle$ includes a wormhole correlating $Y(T)$ and $Y(T)$, and $\langle Y(T)^2 \big(Y(T)^*\big)^2 \rangle$ includes wormholes which pair each copy of $Y(T)$ with a copy of $Y(T)^*$. The double cone saddle point has a compact zero mode, labeled $\delta t$, which appears as a parameter of the $LR$ fields, but not the $LL$ and $RR$ fields. 

The double cone saddle point has an analog in gravity. The geometry is a periodic identification of the two-sided eternal black hole, and the zero mode $\delta t$ corresponds to a relative time shift between the $L$ and $R$ boundaries.

There are also ``disk'' contributions to averages of $Y(T)$, but these contributions decay and can be ignored for sufficiently long times $T$. For simplicity, we will restrict our attention to these timescales. Then we can approximate averages of $Y(T)$ and $Y(T)^*$ just using combinations of double cone wormhole saddle points.\footnote{At very long times there will be additional contributions that compete with the double cone.}

Our proposed semiclassical formula for $\Z(Y_L(T) ) \equiv \Z(Y_L)$ includes solely the contribution of a half-wormhole, corresponding to the $11$ component of the double cone solution for $\langle \Z(Y_L)_1 \Z(Y_L)_2^*\rangle$. Note that because the $11$ (and $22$) components of the double cone saddle point are independent of the zero mode $\delta t$, we should not include an integral over $\delta t$ in the contribution of the half-wormhole. Multiplying by the complex conjugate, we write the schematic formula
\be\label{formulaone}
\Z(Y_L)\Z( Y_R^*) \supset e^{-I_{fl}^{(L)}(\text{Half-wormhole})}e^{-I_{fl}^{(R)}(\text{Half-wormhole})}.
\ee
The formula for $\Z( Y_L(T) Y_R(T)^*)\equiv \Z(Y_L Y_R^*)$ is a sum of contributions from the double cone, including an integral over the zero mode $\delta t$, and a pair of half-wormholes, which for reasons which will be clear in a moment we refer to as ``linked half-wormholes''. Schematically,
\be\label{formulatwo}
\Z(Y_L Y_R^*) \supset e^{-I_{fl}^{(LR)}(\text{Wormhole})}+e^{-I_{fl}^{(LR)}(\text{Linked half-wormholes})}.
\ee
For simplicity we do not explicitly write the integral over $\delta t$. The $LL$ and $RR$ components of the pair of half-wormholes in both (\ref{formulaone}) and the linked half-wormholes in (\ref{formulatwo}) are identical. The $LR$ variables for the linked half-wormholes in (\ref{formulatwo}) are set to zero.

We represent the approximate equality between these two semiclassical expressions with the following figure:
\be\label{SFFfactorizationcondition}
\includegraphics[valign = c, width = .9\textwidth]{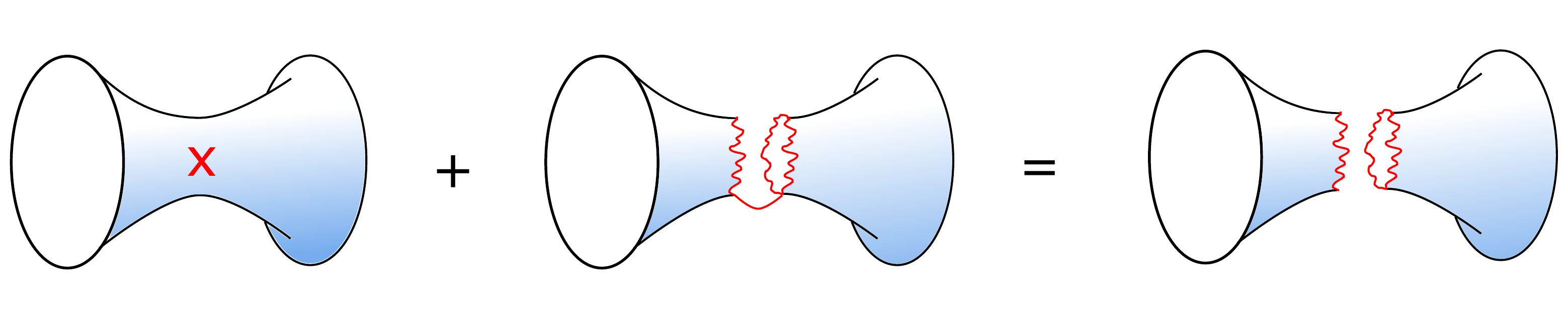}
\ee
These pictures represent saddle points of the SYK path integral, associated to the sketched bulk topology by the pattern of correlation. As the wormhole contribution is self-averaging, we have depicted it with a small red ``x'' to indicate the small amount of randomness. The half-wormhole contributions are not self-averaging, so we have depicted them as ``half'' of a wormhole with a jagged red boundary to indicate the large amount of randomness. We have included a red line linking the pair of half-wormholes on the LHS, to remind us that the $LR$ collective fields are present, but set to zero, distinguishing this contribution from the unlinked pair of half wormholes on the RHS.\footnote{The situation we describe here is similar to the one arrived at in an analysis of the Marolf-Maxfield and JT gravity ensembles in \cite{SSYtoappear,Saadtalk}, so we use similar pictures to illustrate it.}

Perhaps surprisingly, the LHS includes self-averaging and non-self-averaging contributions, while the RHS includes only non-self-averaging terms. One might have instead expected to find an identification just between different half-wormhole contributions.

At the level of the $G,\Sigma$ configurations, such an identification may seem sensible, as the $LL$ and $RR$ components of the two types of half-wormhole contributions in (\ref{SFFfactorizationcondition}) are identical. The difference is simply that the $LR$ variables are set to zero for the linked half-wormholes on the LHS, while they were not introduced in the first place on the LHS.

However, setting the $LR$ variables to zero is very different than not including them in the integral. To see this, we examine the relationship between the actions $I_{fl}^{(LR)}$ and $I_{fl}^{(L)}$, $I_{fl}^{(R)}$,
\be\label{relationshiplrfields}
\int \D G_{LR} \D \Sigma_{LR} \; e^{-I_{fl}^{(LR)} } = e^{-I_{fl}^{(L)}}e^{-I_{fl}^{(R)}}.
\ee
We see that $I_{fl}^{(LR)}\big|_{\text{LR variables}\rightarrow 0}\neq I_{fl}^{(L)}+ I_{fl}^{(R)}$, so that the actions of the two half-wormhole contributions are different. One way to understand this in more detail is as follows. We can represent the right hand side of (\ref{relationshiplrfields}) as an integral over the SYK fermions $\psi^L_i(\tau)$ and $\psi^R_i(\tau')$, generalizing (\ref{intterms}). We then insert a factor of one in the integral in the form\footnote{The coefficient of the $G_{LR}^q$ term is chosen so that averaging over the $J$ couplings with the distribution (\ref{Jdist}) reproduces the conventional averaged $G,\Sigma$ integral.}
\begin{align}
1 \propto \int \D G_{LR} \D \Sigma_{LR} \;\exp\bigg\{ -\frac{N}{2}& \int \hspace{-5pt}\int \; \Sigma_{LR} \big(G_{LR}- \frac{1}{N}\psi^L_i\psi^R_i\big) - \frac{(-1)^{q/2}}{q}\bigg( G_{LR}^q - \big(\frac{1}{N}\psi_i^L \psi_i^R \big)^q \bigg)\bigg\}.
\end{align}
For simplicity we have suppressed the time arguments $\tau,\tau'$ of the variables in this expression.

If we fix the values of $G_{LR}$ and $\Sigma_{LR}$, instead of performing the full integral, this gives an interaction term for the $L$ and $R$ fermions. In particular, if we set both $G_{LR}$ and $\Sigma_{LR}$ to zero, we are left with an interaction like $(\psi^L \psi^R)^q$. In the case we take $Z_L$ and $Z_R$ to be $z$, we can see this residual interaction appearing in the last term of (\ref{identity}).

The lesson is that to see the equality between the two sides of (\ref{relationshiplrfields}) we must do the full integral over the $LR$ variables, rather than set the $LR$ variables to zero. However, if we evaluate the integral semiclassically, we find a \textit{sum} over saddle points with fixed $LR$ variables which \textit{approximately} factorizes.

In the case that the $LL$ and $RR$ fields implicit in (\ref{relationshiplrfields}) are set to half-wormhole values, we identify the two terms in (\ref{formulaone}) as the contributions of two saddle points in the integral over the $LR$ fields. This sum over saddle points is what is needed to restore factorization.
\begin{align}
\int \D G_{LR} \D \Sigma_{LR} \; e^{-I_{fl}^{(LR)}}\bigg|_{\parbox{6em}{{\scriptsize LL and RR fields $\rightarrow$ Half-wormholes}}} \ &\approx \ e^{-I_{fl}^{(LR)}(\text{Wormhole})}+e^{-I_{fl}^{(LR)}(\text{Linked half-wormholes})}
\cr
& \approx \ e^{-I_{fl}^{(L)}(\text{Half-wormhole})}e^{-I_{fl}^{(R)}(\text{Half-wormhole})}.
\end{align}
To summarize, we can think of this as follows: equation (\ref{relationshiplrfields}) tells us that the product of half-wormhole contributions in (\ref{formulatwo}) contains the whole integral over the $LR$ variables. This integral includes both the wormhole and linked half-wormholes, so \textit{the wormhole is contained in the product of two unlinked half-wormholes}.

\section{ Discussion}\label{sectiondiscussion}

In this paper we have studied simple SYK-like models with \textit{fixed} couplings, where factorization from the boundary point of view is manifest.
In the collective field description we have found that the wormhole saddle persists, and that new saddle points exist as well -- half-wormholes.  The combined contribution of these two types of saddles restores factorization in a semiclassical description.  The half-wormhole contribution depends strongly on the particular choice of microscopic fermion couplings, while that of the wormholes only depends weakly.  This non-self-averaging contribution of the half-wormhole explains the strong fluctuations present in such factorizing quantities.  After averaging the half-wormholes disappear, leaving only the non-factorizing wormhole.

 The most important question raised by these findings is what, if any, analogs of such contributions exist in standard holographic theories like Super-Yang-Mills.   Such structures would have to reflect details of the microscopic dynamics of these theories.   Is the ``fuzzball" story relevant here \cite{Mathur:2009hf,Mathur:2014nja}?
 
Assuming these structures do play a role, some more specific questions along these lines would include:
\begin{itemize}
\item Is the half-wormhole geometrically half of a wormhole?  What is its relation to horizon dynamics?
\item What distinguishes linked half-wormholes from unlinked pairs of half-wormholes?
\item What is the analog of $\Phi(\sigma)$ in such a situation? Is it an effective description of some underlying microscopic dynamics?
 \item Is the fact that the wormhole and disk are not {\it exactly} self-averaging significant? is this some perturbative hint about the half-wormholes? What would be the bulk analog?
\end{itemize}

Of course  there are many other questions, including:
\begin{itemize}
\item How can we confirm the picture presented in section \ref{sectiongeneral} for full SYK?   Are there circumstances where we should expect these arguments to fail?
\item Do these ideas apply to tensor models,\footnote{For a review, see \cite{Klebanov:2018fzb} .} which can be viewed as particular, highly structured members of an SYK ensemble?     Is the sparse SYK model \cite{Xu:2020shn} a useful stepping-stone?
\item How do these ideas connect to other approaches to the factorization problem?  
\begin{itemize}
\item Third-quantized ``universe field theory"  approaches \cite{Coleman:1988cy,Giddings:1988cx,Marolf:2020xie} involve a wavefunction in a Hilbert space describing arbitrarily many copies of the boundary theory.  The effective model  presented in \cite{SSYtoappear,Saadtalk} connects that  description to one close to that  presented here.
\item Eberhardt \cite{Eberhardt:2021jvj} employed the localization of  tensionless string worldsheets and a higher genus version of boundary modular invariance and its connection to bulk geometry  to argue for the equivalence of a wormhole partition function  to a factorized one.    Marolf and Maxfield \cite{Marolf:2020xie} uncovered a ``quantum equivalence" due to null vectors in the third-quantized Hilbert space.   Here we employ the equivalence of different ``bulk" descriptions, i.e., different choices of collective field representation, to establish factorization.  Are these ideas at all related?
\end{itemize}
\item The collective fields $G, \Sigma$ seem to be the analog of closed string degrees of freedom in the bulk.   Equation \eqref{z2deriv} indicates that the contribution of a  highly nonclassical region of their configuration space -- $G$ near 0, $\Sigma$ on its defining contour -- provides an alternate description of the boundary fermion degrees of freedom.  Is there an analog of this in other holographic systems?
\end{itemize}

\section*{Acknowledgements} 

We thank Nati Seiberg, Milind Shyani, David Simmons-Duffin, and Zhenbin Yang for illuminating discussions. PS was supported by the Marvin L. Goldberger Membership and W. M. Keck Foundation Fund at the Institute for Advanced Study. SS and SY were supported in part by NSF grant PHY-1720397. DS was supported in part by DOE grant DE-SC0021085. 
\appendix

\section{Glossary of $Z$ symbols}\label{app:glossary}

\begin{itemize}
\item[$Z$:]   A partition function or a product of partition functions of a general SYK type system.  We sometimes use $Z_L$ to denote a single copy, and $Z_L Z_R$ to denote a product of two copies. Special cases include the following
\begin{itemize}[leftmargin=+.5in]
\item[$z$:] The partition function of an SYK model with one time point.  (First defined \hyperref[defz]{{\bf here}}.)
\item[$\zeta(\mu)$:] The partition function of an SYK model with two time points and a coupling of strength $\mu$ between them. $\zeta(0) = z^2$.  (First defined \hyperref[defzeta]{{\bf here}}.)
\item[$Y(T)$:] A microcanonical version of the analytically continued thermal partition function $Z(\beta+\i T)$ of the full SYK model. See \cite{Saad:2018bqo}.
\end{itemize}
\item[$\Z(\cdot)$:] The collective field representation that includes all off-diagonal collective fields. So $\Z(Z_L)$ includes only $LL$ fields and $\Z(Z_L)\Z(Z_R)$ includes only $LL$ and $RR$ fields, but $\Z(Z_LZ_R)$ includes the full matrix of $LL$, $RR$ and $LR$ fields. (First defined \hyperref[gsigmanotfactorized]{{\bf here}}.)
\item[$\Z$:] A general-purpose notation for $\Z(\text{something})$. 
\item[$\Zs$:]   An approximation to $\Z(\text{something})$ consisting of a sum over saddles, including wormholes and half-wormholes.
(First defined \hyperref[zdefthree]{{\bf here}}.)
\end{itemize}

\section{Properties of $z$ for  \texorpdfstring{$q = 2$}{q=2}}\label{appendixqequaltwo}
\subsection{Overview}
In this Appendix we discuss some properties of the single time point model for $q=2$.  This case differs from the generic $q>2$ case analyzed in  \hyperref[sectiononepoint]{{\bf section two}} in several ways:
\begin{enumerate}
\item The $q =2$ wormhole saddle in $z^2$ is not within the self-averaging region.  
\item The $q =2$ half-wormhole saddle is not isolated, as it is for $q>2$,  but is part of a smooth manifold which joins onto the wormhole saddle. This is due to an enhanced continuous symmetry of the collective field action.
\item As a consequence, the fluctuations of $z$ in the $J$ ensemble are much larger than for $q>2$, and grow with $N$.
\item The model is numerically tractable, and we can illustrate these findings with numerical results for reasonably large $N$.  

\end{enumerate}
\subsection{Calculation of  $\lr \Phi^2(\sigma) \rr$}
We first study the statistical properties of  $ \Phi(\sigma)$  \ref{intterms} for $q=2$. More concretely, we calculate its second moment $\lr \Phi^2(\sigma) \rr$ by introducing collective fields as in \ref{Phisquared}. The  $g_{AB}$  integral is now Gaussian so we can perform it exactly, finding
\begin{equation}
\left\langle\Phi^{2}(\sigma)\right\rangle=\int_{\mathbb{R}} \frac{\d^{4} \sigma_{A B}}{(2 \pi / N)^{2}} \exp \left\{N\left[\log \left(\sigma^{2}+\sigma_{\L\RR} \sigma_{\R\LL}-\sigma_{\L\LL} \sigma_{\R\RR}\right)-\frac{1}{2} \sigma_{A B}^{2}\right]\right\},
\end{equation}
in which $(A, B) \in\{(\L,\LL),(\R,\RR),(\L,\RR),(\R,\LL)\}$

This can be calculated by saddle point at large $N$. 
For  $|\sigma|>1$ a saddle point with all $\sigma_{AB}=0$ dominates.  For $|\sigma|<1$   a manifold of saddle points, a consequence of the enhanced symmetry of the problem, dominates. This manifold is a circle, described by:
\begin{equation}
\begin{array}{ll}
\text { circle 1: } & \sigma^{2}+x^{2}+y^{2}=1, \quad \sigma_{\L\RR}=\sigma_{\R\LL}=x, \quad \sigma_{\L\LL}=-\sigma_{\R\RR}=y \\
\text { circle 2: } & \sigma^{2}-x^{2}-y^{2}=-1, \quad \sigma_{\L\RR}=-\sigma_{\R\LL}=x, \quad \sigma_{\L\LL}=\sigma_{\R\RR}=y .
\end{array}
\end{equation}
Integrating over the collective coordinate we obtain:
\begin{equation}
\left\langle\Phi(\sigma)^{2}\right\rangle \approx\left\{\begin{array}{ll}
\sqrt{\frac{\pi N}{2}}\left[e^{N\left(\sigma^{2}-1\right)}+(-1)^{N} e^{N\left(-\sigma^{2}-1\right)}\right] & |\sigma|<1 \\
\frac{\sigma^{4}}{\sigma^{4}-1} e^{2 N \log \sigma} & |\sigma|>1 \\
\text { region shrinks as $N\to \infty$ } & |\sigma| \approx 1
\end{array}\right.
\end{equation}
The factor of  $\sqrt{N}$ is the inverse coupling constant familiar from collective coordinate calculations. Unfortunately neither saddle point approximation gives an accurate answer for $\langle \Phi(1)^2\rangle$, which is the value relevant for determining whether the wormhole saddle at $\sigma = 1$ is self-averaging. However, we can use the exact answer \ref{exactphi2}, which simplifies for $q = 2$ thanks to the identity
\be
\sum_{n_1+n_2 = m} \frac{(2n_1)!(2n_2)!}{(n_1!)^2(n_2!)^2} = 4^m.
\ee
Using this, one finds
\be
\langle \Phi(\sigma)^2\rangle = \frac{N!}{N^N}\sum_{n_1=0}^{N/2} \frac{(N\sigma)^{2n_1}}{(2n_1)!}, \hspace{20pt} (q = 2).
\ee
For large $N$ and $\sigma =1$, the sum is dominated by values of $n_1$ within $O(\sqrt{N})$ of the maximal value $N/2$. Using Stirling's approximation and approximating the sum as an integral, we find $\langle \Phi(1)^2\rangle  = \sqrt{2\pi N}/4 + O(1)$. Note that this is much larger than $\langle \Phi(1)\rangle^2 = 1$. This implies that $\Phi(\sigma)$ at the wormhole saddle $\sigma = 1$ is not self-averaging ($\Phi(\sigma)$ becomes self-averaging only for $|\sigma|\gg 1$).

\subsection{Numerics for the non-averaged system}
For $q=2$ numerics for $\Phi(\sigma)$ are tractable for a fixed set of $J_{ij}$ and reasonably large $N$.   We evaluate
\begin{equation}
\Phi(\sigma)=\int \mathrm{d}^{2 N} \psi \exp \left\{ \sigma \psi_{i}^{\L} \psi_{i}^{\R}+\mathrm{i} J_{ij}\left(\psi_{ij}^{L}+\psi_{ij}^{R}\right)-\frac{N}{2}\left(\frac{1}{N} \psi_{i}^{L} \psi_{i}^{\R}\right)^{2}\right\}
\end{equation}
by introducing a collective field and writing
\begin{equation}
\Phi(\sigma) =\frac{\sqrt{N}}{\sqrt{2 \pi} } \int_{-\infty}^{\infty} \mathrm{d} s \phi(\sigma+\mathrm{i} s) e^{-\frac{N}{2} s^{2}}.
\end{equation}
Here $\phi(\sigma+\i s)$ is the efficiently computable quantity (a Pfaffian):
\begin{equation}
\phi(\sigma)=\int \mathrm{d}^{2N} \psi \ e^{\sigma \sum_{i} \psi_{i}^\L \psi_{i}^\R+\mathrm{i} J_{i j}\left(\psi_{i j}^\L+\psi_{i j}^\R\right)}.
\end{equation}

\begin{figure}[H]
\begin{center}
\includegraphics[width = .6\textwidth]{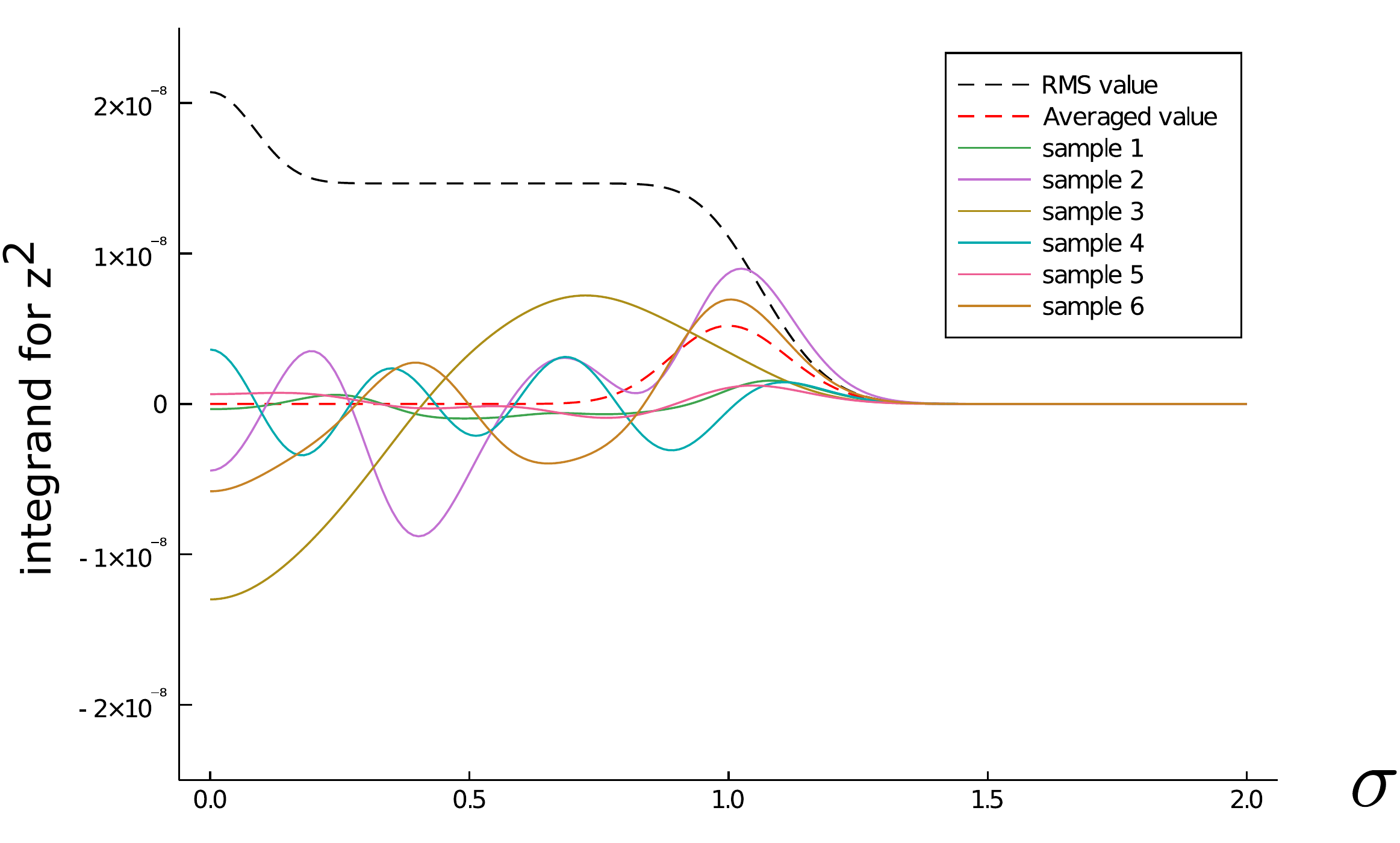}
\caption{{\sf We plot six different samples of the integrand for $z^2$, namely $\Phi(\sigma) \Psi(\sigma)$. Solid lines with different colors denote different samples. The black dashed line is the exact RMS value of the integrand using (\ref{exactphi2}), and the red dashed line is the exact averaged value, using (\ref{averagedPhi}). Here $N=40$, $q=2$. The wormhole saddle for $\langle z^2\rangle$ is at a value $\sigma = 1$, which is not within the self-averaging region.}}\label{numphi}
\end{center} 
\vspace{-2em}
\end{figure}

In figure \ref{numphi}, we plot the $z^2$ integrand, given by $\Phi(\sigma) \Psi(\sigma) = \sqrt{\frac{N}{2\pi}}\Phi(\sigma) e^{-\frac{N}{2}\sigma^2}$, for several different random samples of the couplings $J_{ij}$, and for $N = 40$. We see clearly that the entire region in which the integrand is large is non-self-averaging. This is unlike the case with $q > 2$, where there would be two competitive regions, one self-averaging and the other not. Note also that the values in this region for the six samples illustrated  are typically substantially smaller than the RMS value.  In fact  rare $J_{ij}$ choices make a large contribution to the RMS value. We will see further evidence for this in the computation of higher moments.

\subsection{  $\langle z^k \rangle$ for arbitrary $k$}\label{subsectionq2arbitraryk}

The Gaussian nature of the $q=2$ problem enlarges the symmetry of the collective field  representation of $\langle z^k \rangle$.  For $q>2$ this problem has a discrete $S_k$ permutation symmetry among the $k$ time points.   For $q=2$ this is enhanced to an $O(k)$ continuous symmetry, where $O(k)$ is the orthogonal group.  After integrating out the Gaussian $g_{ab}$ fields we can write the result as a matrix integral over the independent components of a real antisymmetric $k \times k$ matrix $\sigma_{ab}$
\begin{equation}\label{zkintegral}
\langle z^k \rangle  = \left(\frac{N}{2\pi}\right)^{k(k-1)/4} \int_{-\infty}^\infty \d \sigma_{ab}\text{Pf}(\sigma)^N \exp{(\frac{N}{4} {\rm Tr}~ \sigma^2)}.
\end{equation}
Here the integral is over the independent components $a < b$. This integral has a manifest $O(k)$ symmetry, where $\sigma \rightarrow O \sigma O^T$. This can be used to bring $\sigma$ to block diagonal form with blocks
\be\label{evaluematrixform}
\left(\begin{array}{cc}0 & \lambda_i \\ -\lambda_i & 0\end{array}\right), \hspace{20pt} \lambda_i > 0.
\ee
So the integral can be reduced to an integral over the $O(k)$-invariant eigenvalues $\lambda_i$, with an appropriate measure, times the volume of the $O(k)$ divided by the volume of the subgroup that leaves the block-diagonal form invariant. Explicitly,
\be
\prod_{1\le a < b \le k}\int_{-\infty}^\infty \d \sigma_{ab} = \text{vol}\left(\frac{O(k)}{SO(2)^{k/2}S_{k/2}}\right)\prod_{1\le i\le k/2}\int_{0}^\infty \d\lambda_i \prod_{1\le i<j\le k/2}(\lambda_i^2-\lambda_j^2)^2.
\ee
With the normalizations that were used in evaluating the measure, we have
\be
\text{vol}(SO(2)) = 2\pi, \hspace{20pt} \text{vol}(O(k)) = \prod_{m = 1}^k \frac{2 \pi^{m/2}}{\Gamma(m/2)}
\ee
and of course $\text{vol}(S_{k/2}) = (k/2)!$.

Now, the integral we want is (assuming $N$ is even and $k$ is even)
\begin{align}\label{sigmaevalueint}
\langle z^k\rangle 
&= \left(\frac{N}{2\pi}\right)^{k(k-1)/4}\text{vol}\left(\frac{O(k)}{SO(2)^{k/2}S_{k/2}}\right)\prod_{i=1}^{k/2}\int_{0}^\infty \d\lambda_i \left[\prod_{1\le i<j\le k/2}(\lambda_i^2-\lambda_j^2)^2\right] e^{N\sum_{i=1}^{k/2} (\log(\lambda_i) - \frac{1}{2}\lambda_i^2)}.
\end{align}
For $N \gg k$, the eigenvalues will all be close to the saddle point value $\lambda = 1$ of the potential in the last factor.  So expand around this point, writing $\lambda_i = 1 + x_i$. Then we have

\begin{align}\label{larg}
\langle z^k\rangle &\approx \left(\frac{N}{2\pi}\right)^{k(k-1)/4}\text{vol}\left(\frac{O(k)}{SO(2)^{k/2}S_{k/2}}\right)\prod_{i=1}^{k/2}\int_{-\infty}^\infty \d x_i \left[\prod_{1\le i<j\le k/2}4(x_i-x_j)^2\right] e^{N\sum_{i=1}^{k/2} (-\frac{1}{2}-x_i^2)}\notag \\
&= \left(\frac{N}{2\pi}\right)^{k(k-1)/4}\text{vol}\left(\frac{O(k)}{SO(2)^{k/2}S_{k/2}}\right)4^{k(k-2)/8}\left(\frac{\pi}{N}\right)^{k^2/8}\frac{\text{vol}(U(1)^{k/2}S_{k/2})}{\text{vol }U(k/2)}e^{-kN/4}.
\end{align}
In going to the second line, we interpreted the integral over $x_i$ as the integral over the eigenvalues of an ordinary $k/2\times k/2$ Hermitian matrix integral with a quadratic potential. This integral can be done easily in terms of the matrix elements, and going from the eigenvalues to the matrix elements introduced some group volume factors as in the discussion above. (We are using a normalization where the volume of $U(1)$ is $2\pi$.) The expression can be simplified, and we find\footnote{Related formulas in the Hermitian case and their connection to the Riemann $\zeta$ function are discussed in \cite{2000CMaPh.214..111B}.}
\be\label{kllNans}
\frac{\langle z^k\rangle}{\langle z^2\rangle^{k/2}} = \frac{1}{2^{k/2}}\left(\frac{N}{\pi}\right)^{k(k-2)/8}\frac{\text{vol } O(k)}{\text{vol }U(k/2)} = \left(\frac{N}{2}\right)^{k(k-2)/8}\prod_{m = 1}^{k/2}\frac{\sqrt{\pi}}{\Gamma(m-\frac{1}{2})} \hspace{20pt} (k \ll N).
\ee

Let's try to understand \eqref{kllNans} better.   For $k \ll N$ the first factor, $N^{k(k-2)/8}$, dominates.  The moments are very large, and grow rapidly with $k$.  This indicates a very broad distribution for $z$ with long tails.\footnote{The leading $N^{k(k-2)/8}$ dependence describes the moments of a log normal distribution with variance $\sim \log{N}$.} Such a broad distribution is consistent with  the above  numerical results which indicate that  rare configurations of couplings play an important role in $q=2$.
For $k = 4$ this large result is in sharp contrast to the Gaussian value of $3$ found from the discrete saddle points for $q>2$.

 As we already have seen in the computation of $\langle \Phi(\sigma)^2 \rangle$ above such positive powers of $N$ come from the collective coordinate integral over the  saddle point manifold.   For general $k$ the saddle point manifold  corresponds to the orbit of a  single wormhole pairing saddle  under the action of  $O(k)$.       This space is just the quotient $O(k)/G$ where $G$ is the subgroup of $O(k)$ that leaves the wormhole saddle fixed. The wormhole saddle matrix is of the form \eqref{evaluematrixform} with all $\lambda_i$ equal to the saddle point value $\lambda_s = 1$.  Call this matrix  $\lambda_s \Omega$.    Invariance of the saddle means $O^{\text{T}} \Omega O = \Omega$.  So $O$ must be a symplectic matrix as well.  The intersection of $O(k)$ with $Sp(2k)$ is just $U(k/2)$.\footnote{This is referred to as the ``2 out of 3 property" and follows from the embedding of ${\bold C}^{k/2}$ in ${\bold R}^{k}$ using $\Omega$ as a complex structure.  See for example \cite{1497300}.}    So the manifold of saddles is $O(k)/U(k/2)$.    The dimension of this space is $k(k-1)/2 - (k/2)^2 = k(k-2)/4$.   We get a factor of $N^{1/2}$ for every collective coordinate giving an expected factor of $N^{k(k-2)/8}$, which agrees with \eqref{kllNans}.

The first factor in \eqref{kllNans} controls its behavior when $k$ is fixed (but $\gg 1$) and $N \to \infty$.   But when $k$ increases the second factor  becomes important.  Using Stirling's formula for $k \gg 1$ we find \eqref{kllNans} behaves like

\be\label{klessthannapprox}
\frac{\langle z^k\rangle}{\langle z^2\rangle^{k/2}} \sim
(\frac{N}{k})^{k^2/8}
\ee
Equations \eqref{kllNans} and \eqref{klessthannapprox} are only valid for $k \ll N$ but we can see qualitatively that when $k$ becomes a finite (if small) fraction of $N$ the behavior of the moments changes.  

 We can understand  the origin of this scale by looking at the balance of terms in the matrix integral \eqref{sigmaevalueint}.   Putting everything in the exponential we see that the ``potential" terms are of order $Nk$ and the Vandermonde term is of order $k^2$.   When $k \ll N$ the effect of the Vandermonde is small and the eigenvalues sit at the saddle point of the potential.  This describes the wormhole saddle.    But when $k \sim N$ the Vandemonde is important and the eigenvalues are pushed away from the wormhole saddle.

 Let's now try to understand the behavior in the opposite limit $k \gg N$.  It turns out to be more useful to return to the original SYK variables using the defining equation \eqref{z2defn}, generalized to arbitrary $k$.   We can integrate out the fermions immediately to get an expression of the form
 \be\label{jintegral}
 \langle z^k \rangle = \left(\frac{N}{2\pi}\right)^{N(N-1)/4}\int  d J_{ab} ~ \text{Pf}(J)^k ~\exp{( \frac{N}{4}\text{tr} J^2)}~.
 \ee
  This is a matrix integral of the same type as \eqref{zkintegral} and can be analyzed the same way.\footnote{This equivalence is a version of Brezin-Hikami duality \cite{Brezin:2007wv}, is closely related to the color-flavor transformation \cite{1998chao.dyn.10016Z}, and also appears in the low dimensional open-closed string correspondence \cite{Maldacena:2004sn,Gaiotto:2003yb}.}    Note that here  $J$ is an antisymmetric $N \times N$  matrix whose size is fixed at $N$ but the power of its Pfaffian varies with $k$.        We rewrite \eqref{jintegral} in terms of the eigenvalues of $J$, which we denote
  $\mu_i, ~ i = 1 \ldots N/2$. 
  The analog of \eqref{sigmaevalueint} is
 
  \begin{align}\langle z^k\rangle &= \left(\frac{N}{2\pi}\right)^{N(N-1)/4}\prod_{1\le a < b \le N}\int_{-\infty}^\infty \hspace{-1em} d J_{ab}~ \text{Pf}(J)^k e^{-\frac{N}{2}\sum_{1\le a < b \le N}J_{ab}^2}\\
&= \left(\frac{N}{2\pi}\right)^{N(N-1)/4}\hspace{-1em}\text{vol}\left(\frac{O(N)}{SO(2)^{N/2}S_{N/2}}\right)\prod_{i=1}^{N/2}\int_{0}^\infty \hspace{-1em} d\mu_i \left[\prod_{1\le i<j\le N/2}(\mu_i^2-\mu_j^2)^2\right] e^{\sum_{i=1}^{N/2} (k\log(\mu_i) - \frac{N}{2}\mu_i^2)}\label{muintegral}
\end{align}
Extremizing the potential in the last factor of  \eqref{muintegral} we find a saddle point at $\mu_s = \sqrt{k/N}$.   Fluctuations in $\mu_i$ are small if $N \gg 1$ so as before we expand $\mu_i = \mu_s + x_i$.  But now the number of eigenvalues is $N$, independent of $k$, and the coefficient of $x_i^2$ is $N$, $k$ independent as well.  So the integral over the eigenvalue fluctuations does not contribute any nontrivial $k$ dependence, apart from an overall scaling.   Evaluating the potential (and the Vandermonde scaling) at $\mu_s$ we find the result
\be\label{momentskggn}
\frac{\langle z^k\rangle}{\langle z^2\rangle^{k/2}} \approx \frac{1}{2^{\frac{N+k}{4}}}\frac{\text{vol }O(N)}{\text{vol }U(N/2)}\left(\frac{k}{N}\right)^{\frac{Nk}{4}}\left(\frac{k}{\pi}\right)^{\frac{N(N-2)}{8}} \hspace{20pt} (k\gg N \gg 1).
\ee

 The moments again grow rapidly with $k$ indicating a long tail in the  $z$ distribution for the largest $z$ values.  Again we see a change in behavior when $k \sim N$ (although here $N$ must be a small fraction of $k$ to remain in the domain of validity of \eqref{momentskggn}). The last factor is again due to the collective coordinate integral but here is a subleading effect.
 
 The change of behavior from \eqref{klessthannapprox} when $k \gg N$ is due to the localization of $J$ to very special matrices.  The high power of the Pfaffian selects matrices that maximize it subject to the $J$ Gaussian weight.  The  eigenvalues $\mu_i$ are all equal and very large, $(\frac{k}{N})^{\frac{1}{2}}$.   This is highly atypical from the point of view of the original Gaussian ensemble for $J$, where the typical size of one $J_{ab}$ is of order $ (\frac{1}{N})^{\frac{1}{2}}$, and has Gaussian fluctuations.

 We can compare these results to the $q > 2$ case. For $q>2$ the leading corrections to the saddle point analysis are of order $1/N^{q-2}$ and at this order there are of order $k^2$ terms.  So we estimate that the leading order corrections are of order $k^2/N^{(q-2)}$.   This predicts a change in behavior at $k \sim N^{(q-2)/2}$. In the case examined here, $q=2$, this corresponds to the fact that even the low order moments are not Gaussian (which is the behavior we expect for $q>2$) due to the presence of the zero mode.  

As we just saw, the $k \sim N$ scale indicates the dominance of certain highly atypical $J$ configurations.  We expect something similar for $q>2$ at large enough $k$.   Rare $J$'s which maximize $z$ will be strongly favored.   We do not know if this occurs at the $k \sim N^{(q-2)/2}$ scale or at parametrically larger values of $k$.  

More generally, we expect the breakdown in the wormhole picture of the moments for arbitrarily high moments to be an indication of this focus on highly atypical ``Hamiltonians" in the SYK ensemble.   The description of the higher moments is simpler in terms of the fixed (order $N$)  ``boundary" variables $\psi_i, J_{i_1 ... i_q}$.

\section{More on the $\Psi$ function}\label{Psiapp}
In this appendix we will give some details on the $\Psi$ function for large $N$. We will focus on the case $q = 4$. The definition is
\be\label{psiappeq1}
\Psi(\sigma) = \int_{-\infty}^\infty \frac{\d g}{2\pi/N} e^{N(-\i \sigma g - \frac{1}{4}g^4)}.
\ee
The saddle point equations have three solutions, and depending on the phase of $\sigma$, the integral may receive contributions from one or two of these saddle points. The analysis of this is similar to the one for the Airy function, and in the complex $\sigma$ plane, one finds the following behavior for large $N$ (up to one-loop prefactors that we will omit in this appendix):
\be\label{stokes}
\includegraphics[width = .5\textwidth,  valign = c]{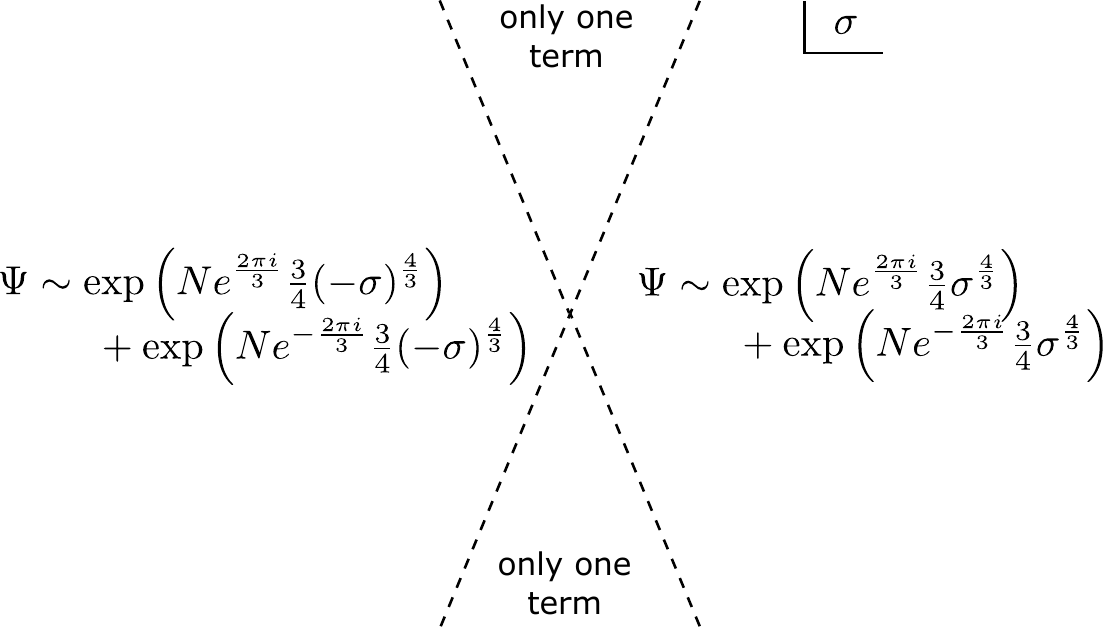}
\ee
The Stokes rays are at angle $e^{\frac{3\pi \i}{8}}$ and reflections across the real and imaginary axes. Note
\be
e^{\frac{2\pi \i}{3}} = -\tfrac{1}{2} + \sqrt{\tfrac{3}{4}} \ \i
\ee
so the function is decaying and oscillating along the real axis.

This behavior can be used to give a simple explanation for why all $q$ (in this case four) saddles contribute to $\langle z^2\rangle$, as an alternative to the higher-dimensional Lefscehtz thimble method. Consider the integral
\be\label{zsappendix}
\langle z^2\rangle = \int_{-\infty}^\infty \d \sigma \sigma^{2N} \Psi(\sigma).
\ee
By inserting the semiclassical approximation (\ref{stokes}), we write this as
\begin{align}
\langle z^2 \rangle &\sim \int_0^\infty \d \sigma \sigma^{2N} \exp\left(N e^{\frac{2\pi \i}{3}}\tfrac{3}{4}\sigma^{\frac{4}{3}}\right)+\int_0^\infty \d \sigma \sigma^{2N} \exp\left(N e^{\frac{-2\pi \i}{3}}\tfrac{3}{4}\sigma^{\frac{4}{3}}\right) \notag \\ &\hspace{20pt}+ \int_{-\infty}^0 \d \sigma \sigma^{2N} \exp\left(N e^{\frac{2\pi \i}{3}}\tfrac{3}{4}(-\sigma)^{\frac{4}{3}}\right) +\int_{-\infty}^0 \d \sigma \sigma^{2N} \exp\left(N e^{\frac{-2\pi \i}{3}}\tfrac{3}{4}(-\sigma)^{\frac{4}{3}}\right)
\end{align}
These four integrals can now be deformed to steepest-descent contours passing through each of the four saddle points for $\langle z^2\rangle$. For example, the first one deforms to a steepest descent contour along the ray $r e^{\frac{\i \pi }{4}}$, passing through a saddle at the point where this ray intersects the unit circle. 

\begin{figure}[h]
\begin{center}
\includegraphics[width=.5\textwidth]{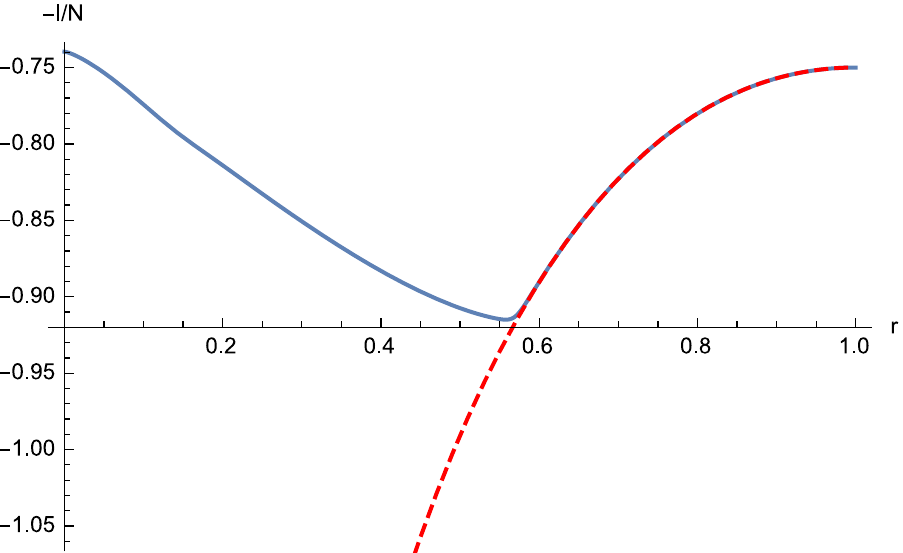}
\caption{{\sf The solid blue curve is the logarithm of the RMS value of the integrand along the ray $\sigma = r e^{\frac{\i \pi}{4}}$, namely $\frac{1}{N}\log(\Phi_{\text{rms}}(e^{\frac{\i\pi}{4}}r)^2) - \frac{3}{4}r^{\frac{4}{3}}$. The dashed red curve is the logarithm of the averaged value of this integrand, namely $\frac{1}{N}\log(\Phi_{\text{mean}}(e^{\frac{\i\pi}{4}}r)^2) - \frac{3}{4}r^{\frac{4}{3}}$. The wormhole saddle point is at $r = 1$, and the half-wormhole is at $r = 0$. For the plot we took $N = 100$ and $q = 4$. (We don't know how to make a plot with samples for $q > 2$ because $\Phi(\sigma)$ seems to be intractable.)}}\label{actionsPlot}
\end{center}
\vspace{-1em}
\end{figure}

One can use a similar strategy in the theory with fixed couplings, after replacing $\sigma^{2N}$ by $\Phi(\sigma)$. In figure \ref{actionsPlot}, we plot the log of the averaged integrand, and the log of the RMS integrand along the ray, $\sigma = e^{\frac{\i\pi}{4}}r$. The boundary of the self-averaging region is clearly visible at $r \approx .56$. Note that in the theory with fixed couplings, the region at $r = 0$ contributes approximately the same as the wormhole saddle at $r = 1$. An apparent problem is that the semiclassical expansion of $\Psi$ breaks down near the origin $\sigma = 0$. However, we can analyze integrals that are peaked in this region using the property
\be\label{normapp}
\int_{-\infty}^\infty \d \sigma \Psi(\sigma) = 1.
\ee
(This is easy to prove using (\ref{psiappeq1}) and interchanging the order of integration.) Because $\Psi(\sigma)$ is rapidly decaying, the important contribution to (\ref{normapp}) comes from within a small distance $N^{-3/4}$ of $\sigma = 0$, so $\Psi(\sigma)$ acts as an approximate delta function. Alternatively, note that although $\sigma = 0$ is a singular point in the semiclassical expansion if $g$ has already been integrated out, if we restore the $g$ variable then there is no problem. Then using the smoothness of $\Phi(\sigma)$ near $\sigma = 0$ (which can be determined from the relation $\langle( \Phi(\sigma) -\Phi(\sigma'))^2\rangle = \langle \Phi^2(\sigma) + \Phi^2(\sigma') - 2 \Phi^2(\sqrt{\sigma \sigma'})\rangle$), we see that the region near $\sigma = 0$ in the integral for $z^2$ (without averaging) can be interpreted as the contribution of a genuine saddle point in the full $g, \sigma$ space.  This is the half-wormhole saddle.

\bibliography{references}

\bibliographystyle{utphys}

\end{document}